\documentclass[twocolumn]{aastex63}
\received{} 
\revised{} 
\accepted{} 
\submitjournal{ApJ}

\shorttitle{Two-component magnetic field along the line of sight to the Perseus Molecular Cloud}
\shortauthors{Doi {\em et al.}}
\graphicspath{{./}{figures/}}

\begin{document}

\title{Two-component Magnetic Field along the Line of Sight to the Perseus Molecular Cloud:\\Contribution of the Foreground Taurus Molecular Cloud}

\correspondingauthor{Yasuo~Doi}
\email{doi@ea.c.u-tokyo.ac.jp}

\author[0000-0001-8746-6548]{Yasuo~Doi}
\affiliation{Department of Earth Science and Astronomy, Graduate School of Arts and Sciences, The University of Tokyo, 3-8-1 Komaba, Meguro, Tokyo 153-8902, Japan}

\author[0000-0003-1853-0184]{Tetsuo Hasegawa}
\affiliation{National Astronomical Observatory of Japan, National Institutes of Natural Sciences, Osawa, Mitaka, Tokyo 181-8588, Japan}

\author[0000-0002-0794-3859]{Pierre Bastien}
\affiliation{Institut de Recherche sur les Exoplan\`etes (iREx), Universit\'e de Montr\'eal, D\'epartement de Physique, C.P. 6128 Succ. Centre-ville, Montr\'eal, QC H3C 3J7, Canada}
\affiliation{Centre de Recherche en Astrophysique du Qu\'ebec (CRAQ), Universit\'e de Montr\'eal, D\'epartement de Physique, C.P. 6128 Succ. Centre-ville, Montr\'eal, QC H3C 3J7, Canada}

\author[0000-0001-8749-1436]{Mehrnoosh Tahani}
\affiliation{Dominion Radio Astrophysical Observatory, Herzberg Astronomy and Astrophysics Research Centre, National Research Council Canada, P. O. Box 248, Penticton, BC V2A 6J9 Canada}

\author{Doris Arzoumanian}
\affiliation{Aix Marseille Univ, CNRS, CNES, LAM, Marseille, France}

\author[0000-0002-0859-0805]{Simon Coud\'e}
\affiliation{SOFIA Science Center, Universities Space Research Association, NASA Ames Research Center, M.S. N232-12, Moffett Field, CA 94035, USA}
\affiliation{Centre de Recherche en Astrophysique du Qu\'ebec (CRAQ), Universit\'e de Montr\'eal, D\'epartement de Physique, C.P. 6128 Succ. Centre-ville, Montr\'eal, QC H3C 3J7, Canada}

\author[0000-0002-6906-0103]{Masafumi Matsumura}
\affiliation{Faculty of Education, \& Center for Educational Development and Support, Kagawa University, Saiwai-cho 1-1, Takamatsu, Kagawa, 760-8522, Japan}

\author[0000-0001-7474-6874]{Sarah Sadavoy}
\affiliation{Department for Physics, Engineering Physics and Astrophysics, Queen's University, Kingston, ON K7L 3N6, Canada}

\author[0000-0002-8975-7573]{Charles L. H. Hull}
\affiliation{National Astronomical Observatory of Japan, NAOJ Chile, Alonso de C\'ordova 3788, Office 61B, 7630422, Vitacura, Santiago, Chile}
\affiliation{Joint ALMA Observatory, Alonso de C\'ordova 3107, Vitacura, Santiago, Chile}
\affiliation{NAOJ Fellow}

\author{Yoshito Shimajiri}
\affiliation{Department of Physics and Astronomy, Graduate School of Science and Engineering, Kagoshima University, 1-21-35 Korimoto, Kagoshima, Kagoshima 890-0065, Japan}
\affiliation{National Astronomical Observatory of Japan, National Institutes of Natural Sciences, Osawa, Mitaka, Tokyo 181-8588, Japan}

\author[0000-0003-0646-8782]{Ray S. Furuya}
\affiliation{Institute of Liberal Arts and Sciences, Tokushima University, Minami Jousanajima-machi 1-1, Tokushima 770-8502, Japan}

\author[0000-0002-6773-459X]{Doug Johnstone}
\affiliation{Herzberg Astronomy and Astrophysics Research Centre, National Research Council of Canada, 5071 West Saanich Rd, Victoria, BC V9E 2E7, Canada}
\affiliation{Department of Physics and Astronomy, University of Victoria, Victoria, BC V8P 5C2, Canada}

\author[0000-0002-6482-8945]{Rene Plume}
\affiliation{Department of Physics \& Astronomy, University of Calgary, 2500 University Dr. NW, Calgary, AB, T2N1N4, Canada}

\author[0000-0003-4366-6518]{Shu-ichiro Inutsuka}
\affiliation{Department of Physics, Graduate School of Science, Nagoya University, Furo-cho, Chikusa-ku, Nagoya 464-8602, Japan}

\author[0000-0003-2815-7774]{Jungmi Kwon}
\affiliation{Department of Astronomy, Graduate School of Science, The University of Tokyo, 7-3-1 Hongo, Bunkyo-ku, Tokyo 113-0033, Japan}

\author[0000-0002-6510-0681]{Motohide Tamura}
\affiliation{Department of Astronomy, Graduate School of Science, The University of Tokyo, 7-3-1 Hongo, Bunkyo-ku, Tokyo 113-0033, Japan}
\affiliation{Astrobiology Center, National Institutes of Natural Sciences, Osawa, Mitaka, Tokyo 181-8588, Japan}
\affiliation{National Astronomical Observatory of Japan, National Institutes of Natural Sciences, Osawa, Mitaka, Tokyo 181-8588, Japan}



\begin{abstract}
Optical stellar polarimetry in the Perseus molecular cloud direction is known to show a fully mixed bi-modal distribution of position angles across the cloud \citep{1990ApJ...359..363G}.
We study the Gaia trigonometric distances to each of these stars and reveal that the two components in position angles trace two different dust clouds along the line of sight.
One component, which shows a polarization angle of $-37.6^\circ \pm 35.2^\circ$ and a higher polarization fraction of $2.0 \pm 1.7$ \%, primarily traces the Perseus molecular cloud at a distance of 300 pc.
The other component, which shows a polarization angle of $+66.8^\circ \pm 19.1^\circ$ and a lower polarization fraction of $0.8 \pm 0.6$ \%, traces a foreground cloud at a distance of 150 pc.
The foreground cloud is faint, with a maximum visual extinction of $\leq 1$ mag.
We identify that foreground cloud as the outer edge of the Taurus molecular cloud.
Between the Perseus and Taurus molecular clouds, we identify a lower-density ellipsoidal dust cavity with a size of 100 -- 160 pc.
This dust cavity locates at $l = 170^\circ,~b = -20^\circ$, and $d = 240$ pc, which corresponds to an H\small{I} shell generally associated with the Per OB2 association.
The two-component polarization signature observed toward the Perseus molecular cloud can therefore be explained by a combination of the plane-of-sky orientations of the magnetic field both at the front and at the back of this dust cavity.
\end{abstract}



\section{Introduction}
\label{sec:introduction}

Interstellar magnetic fields (\emph{B}-fields) are thought to play an important role in the formation of molecular clouds.
As material from the interstellar medium (ISM) flows along field lines onto the nascent clouds, the resulting filamentary structures are expected to extend perpendicular to the interstellar \emph{B}-field (\citealp{2019FrASS...6....5H} for a review).
However, this flow of matter can also influence the morphology of the surrounding \emph{B}-field.
Therefore, the \emph{B}-field structures we observe around molecular clouds are imprinted with their history of an accumulation from the ISM, thus allowing us to trace the processes that led to their formation \citep[e.g.,][]{2018MNRAS.480.2939G}.

The plane-of-sky (POS) component of the \emph{B}-field can be traced by polarimetric observations of optical and near-infrared radiation from stars located behind molecular clouds, as well as thermal continuum emission from interstellar dust particles in those same clouds \citep{1993prpl.conf..279H, 2007JQSRT.106..225L, 2008MNRAS.388..117H, 2009ApJS..182..143M, 2012ARA&A..50...29C}.
Aspherical dust particles irradiated by starlight are charged up by the photoelectric effect, as well as spun up as a result of radiative torques (RATs; \citealp[e.g.,][]{1996ApJ...470..551D,1997ApJ...480..633D,2019ApJ...883..122L}).
These spinning particles are aligned with their rotation axes (i.e., their minor axes) parallel to the \emph{B}-field orientation.
This alignment of the dust particles results in preferential absorption and scattering of the background starlight, which makes the observed starlight polarized in the direction parallel to the POS component of the \emph{B}-field.
In the case of thermal dust emission in sub-mm wavelengths, the preferential emission from the aligned dust particles causes the emission polarized in the direction perpendicular to the POS component of the \emph{B}-field \citep{1966ApJ...144..318S, 1988QJRAS..29..327H, 2015ARA&A..53..501A}.

The Perseus molecular cloud, whose distance is about 300 pc from the sun \citep{2018ApJ...865...73O, 2018ApJ...869...83Z, 2021A&A...645A..55P}, is one of the most active star-forming molecular clouds in the solar vicinity \citep{2008hsf1.book..308B}.
This cloud is associated with a single H\small{I} shell along with the Taurus, Auriga, and California molecular clouds.
It has been suggested that this H\small{I} shell may have been formed by the interstellar bubble around the Perseus OB2 association \citep{2008hsf1.book..308B, 2013ApJ...765..107L, 2019A&A...623A..16S}.

\citet{2020ApJ...899...28D} revealed that the active star-forming region NGC 1333 in the Perseus molecular cloud shows a complex \emph{B}-field structure at spatial scales $< 1$ pc.
In their analysis, the small scale variation of the \emph{B}-field is well explained if the \emph{B}-field is generally perpendicular to the local dense ISM filaments.
Fluctuations in the \emph{B}-field structure at small spatial scales in the Perseus molecular cloud are also found by optical polarimetry (\citealp{1990ApJ...359..363G}; Figure \ref{fig:Goodman}).
The observed polarization angles indicate that POS orientations of the \emph{B}-field ($\theta_\mathrm{star}$) show two distinct populations.
One population has a peak at $\theta_\mathrm{star} \sim -40 ^\circ$ and the other population has a broader peak in position angles at $\theta_\mathrm{star} \sim +70 ^\circ$.
The two populations of vectors show no spatial segregation across the Perseus cloud complex (\citealp{1990ApJ...359..363G}; see Figure \ref{fig:Goodman}).

It is not yet known whether these bi-modal $\theta_\mathrm{star}$ values represent the local variation of \emph{B}-field orientations, differences in the characteristics of the dust particles, or whether that is caused by the superposition of multiple ISM components along the line of sight \citep[LOS;][]{1990ApJ...359..363G, 2002ApJ...574..822M, 2006ApJ...643..932R, 2019ApJ...871L..15G}.
In this paper, we combine optical polarimetry data with Gaia measurements of stellar distances, to determine the multi-layer distribution of the \emph{B}-field in the direction of the Perseus molecular cloud.
With this analysis, we aim to reveal the cause of the bi-modal \emph{B}-field structure observed in the Perseus molecular cloud region, which we find due to a contribution of the foreground Taurus molecular cloud.

This paper is organized as follows.
In Section \ref{sec:data}, we describe the data used for our analysis.
In Section \ref{sec:3D}, we analyze the stellar distances of optical polarimetry and identify contributions from the Perseus and the foreground Taurus molecular clouds.
In Section \ref{sec:discussion}, we discuss the relationship between the H\small{I} shell, which is thought to be associated with the Perseus-Taurus molecular cloud, and the \emph{B}-field distribution we have identified.
We also discuss the alignment between $\theta_\mathrm{star}$ and the Planck-observed \emph{B}-field orientation ($\theta_\mathrm{Planck}$), which gives information on small-scale \emph{B}-field structure below Planck's beam size.
In Section \ref{sec:conclusion}, we summarize the results.

\section{Data}
\label{sec:data}

\subsection{Optical and Near-Infrared Stellar Polarimetry}

We use optical polarimetry of 88 sources in the direction of the Perseus molecular cloud observed by \citet[][Figure \ref{fig:Goodman}]{1990ApJ...359..363G}.
The passband of the observations were centered at 762.5 nm with a bandwidth of 245 nm.

We also use near-infrared (NIR) polarimetry data taken toward NGC 1333 in the Perseus molecular cloud, as shown in the inset of Figure \ref{fig:Goodman}.
We use \emph{K}-band polarimetry by \citet[][14 sources]{1988MNRAS.231..445T} and \emph{R}- and \emph{J}-band polarimetry by \citet[][33 sources]{2011AJ....142...33A}.
The near-infrared $\theta_\mathrm{star}$ are mainly distributed from $\theta_\mathrm{star} \sim -90^\circ$ to $\theta_\mathrm{star} \sim -40^\circ$.

\subsection{Planck Sub-mm Polarimetry}
\label{sec:Planck}

We estimate the \emph{B}-field orientation measured in sub-mm dust emission by using Planck data at 353 GHz \citep{2020A&A...641A..12P}.
We use \verb|HFI_SkyMap_353-psb_2048_R3.01_full.fits| taken from the Planck Legacy Archive, https://pla.esac.esa.int/. 
In our analysis, we set the spatial resolution of the Planck polarimetric data as a $10'$ FWHM Gaussian to achieve good S/Ns.

\subsection{Gaia DR2 Photometry and Trigonometric Distances}
\label{sec:DR2}

We estimate the stellar distances by using Gaia astrometry data (DR2; \citealp{2016A&A...595A...1G, 2018A&A...616A...1G}), and use SIMBAD \citep{2000A&AS..143....9W} positions to cross-match the stars in the Gaia catalog.
We set a search radius of $5\arcsec$ and take the star at the closest position for \emph{R}-band and \emph{J}-band data by \citet{2011AJ....142...33A}.
The identified stars show $M_G = 14$ -- 20 mag, and have good correspondence with the \emph{J}-band magnitude tabulated by \citet{2011AJ....142...33A}.

For relatively old datasets by \citet{1990ApJ...359..363G} and \citet{1988MNRAS.231..445T}, we set a relatively large search radius of $30\arcsec$ and take the brightest star in the searched region, which gives a good cross-match of the catalogued stars as follows.
We find that the stars observed by \citet{1990ApJ...359..363G} show \emph{G}-band magnitude in the Gaia catalog between $M_G = 7$ -- 15 mag.
We exclude three stars that show $M_G = 17$ -- 21 mag from the following analysis for their possible misidentifications.
The stars observed by \citet{1988MNRAS.231..445T} show $M_G = 10$ -- 18 mag. We exclude one star that shows $M_G = 20$ mag from the following analysis because of its non-reliable distance estimation (negative parallax).

The Gaia parallax in DR2 is known to have a systematic bias of -0.03 mas (see \citealp{2015PASP..127..994B}; \citealp{2018AJ....156...58B}; \citealp{2018A&A...616A...2L}; \citealp{2018A&A...616A..17A}; \citealp{2019A&A...621A..48L}; \citealp{2020Ap&SS.365..112M}).
This bias is negligible in our distance evaluation, as the stellar distances in our analysis are less than 1 kpc and thus their parallaxes $\varpi > 1$ mas.
Thus, we did not correct the bias, and estimate the distances of each star by $d = 1000/\varpi$ (pc).

The Renormalised Unit Weight Error (RUWE) is a parameter that is expected to be around 1.0 for sources where the single-star model provides a good fit to the astrometric observations.
A value significantly greater than 1.0 (say, $>1.4$) could indicate that the source is non-single or otherwise problematic for the astrometric solution
(see the Gaia data release documentation\footnote{https://gea.esac.esa.int/archive/documentation/GDR2/}).
We estimate the RUWE for each source following the formulation described in the document ``Re-normalising the astrometric chi-square in Gaia DR2”\footnote{https://www.cosmos.esa.int/web/gaia/public-dpac-documents} and use the data only if whose $\mathrm{RUWE} \leq 1.4$.

As a result, we estimate the distances of 111 sources, including 70 sources from \citet{1990ApJ...359..363G}, 20 sources (\emph{R}-band) and 15 sources (\emph{J}-band) from \citet{2011AJ....142...33A}, and 6 sources from \citet{1988MNRAS.231..445T}.
We summarize the identified data in Table \ref{tab:stellar_data} in Appendix \ref{sec:used_data}.

\section{Results}
\label{sec:3D}

\subsection{Distance to the Clouds that Produce Polarization}
\label{sec:breakpoints}

\begin{figure*}[tp]
\centering
\includegraphics[width=\linewidth]{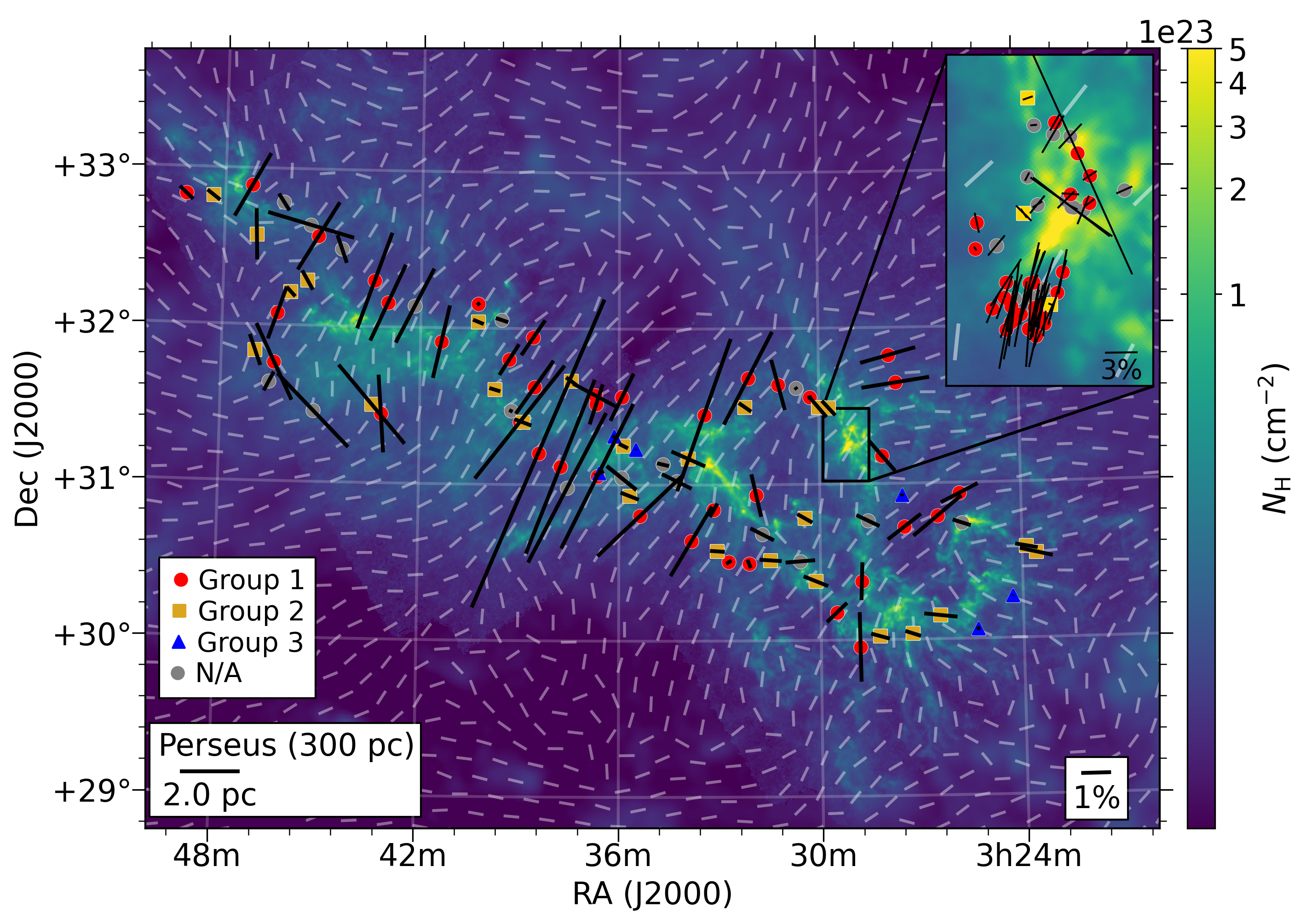}
\caption{Stellar polarimetry data toward the Perseus molecular cloud (black line segments), overlaid on the Planck-observed \emph{B}-field orientation \citep[gray line segments;][]{2020A&A...641A..12P}.
Black line segments in the main panel are the optical polarimetry data by \citet{1990ApJ...359..363G}, and those in the inset are near-infrared polarimetry data by \citet{1988MNRAS.231..445T} and \citet{2011AJ....142...33A} toward NGC 1333.
The length of the black line segments is proportional to the polarization fraction ($P_\mathrm{star}$).
Reference scales of $P_\mathrm{star}$ are shown in the lower right corner of both the main panel and the inset.
Filled circles indicate the position of the stars and are color-coded to indicate stellar distance ranges observed with Gaia.
Group 1 is for distances $d > 300$ pc, Group 2 is for $150 < d < 300$ pc, and Group 3 is for $d < 150$ pc.
N/A is for the data with no reliable distance estimation available.
See Section \ref{sec:breakpoints} for the discussion and the definition of these distance groups.
For the Planck data (gray line segments), the line segments' length has been normalized to show only the orientation of the \emph{B}-field.
The spatial resolution of the Planck data is $10'$.
The background color scale is the hydrogen column density ($N_\mathrm{H}$).
We convert the dust opacity at 353 GHz ($\tau_{353}$), estimated by Herschel and Planck observational data \citep{2016A&A...587A.106Z}, to the hydrogen column density  by $N_\mathrm{H} = 1.6\times 10^{26}~\tau_{353}$ (cm$^{-2}$; \citealp{2014A&A...571A..11P,2015A&A...576A.104P}).
A reference scale of 2 pc, in which we assume the distance to the field is 300 pc, is shown in the lower left corner of the main panel.
}
\label{fig:Goodman}
\end{figure*}

\begin{figure}[tp]
\centering
\includegraphics[width=\linewidth]{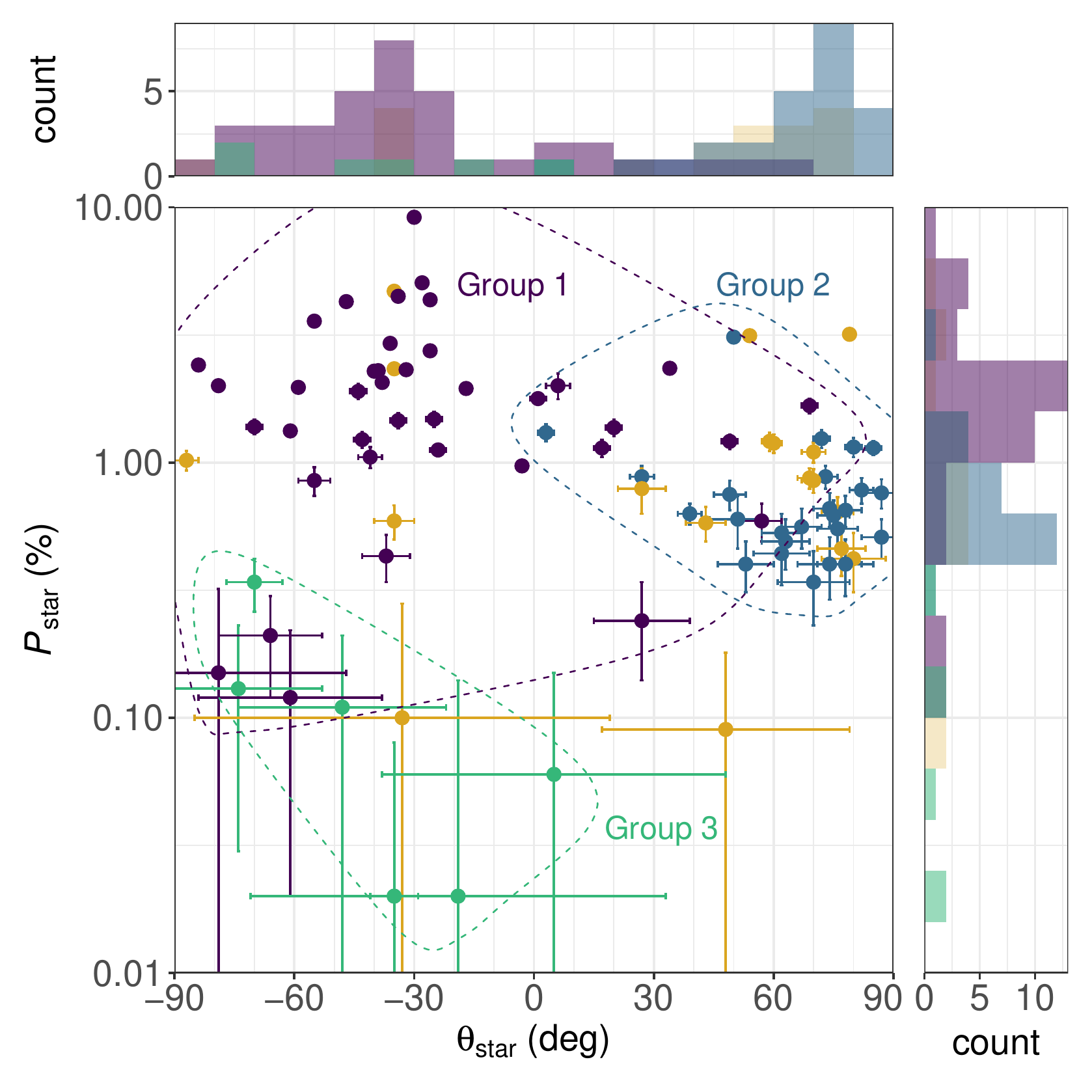}
\caption{
A correlation between $\theta_\mathrm{star}$ and $P_\mathrm{star}$ of the optical polarimetry by \citet{1990ApJ...359..363G}.
Histgrams of the two parameters are also shown on each axis.
Color codes are the stellar distance ranges same as Figure \ref{fig:Goodman}.
Yellow data points are the data with no reliable distance estimation available.
}
\label{fig:GoodmanAngP}
\end{figure}

Figure \ref{fig:Goodman} compares the spatial distribution of the polarimetry data in our analysis with the inferred \emph{B}-field morphology from Planck.
The polarization position angles, $\theta_\mathrm{star}$, show a bi-model distribution, with concentrations at $\theta_\mathrm{star} \sim +70^\circ$ and $\theta_\mathrm{star} \sim -40^\circ$ (measured from North to East).
\citet{1990ApJ...359..363G} found that the two populations in $\theta_\mathrm{star}$ also have distinct polarization fractions ($P_\mathrm{star}$).
We display the relationship between the observed $\theta_\mathrm{star}$ and $P_\mathrm{star}$ in Figure \ref{fig:GoodmanAngP}.
The population of $\theta_\mathrm{star} \sim -40^\circ$ show relatively larger $P_\mathrm{star} \gtrsim 1$\%, while the other population that has $\theta_\mathrm{star} \sim +70^\circ$ show relatively smaller $P_\mathrm{star} \lesssim 1$\%.

\begin{figure}[tp]
\centering
\includegraphics[width=\linewidth]{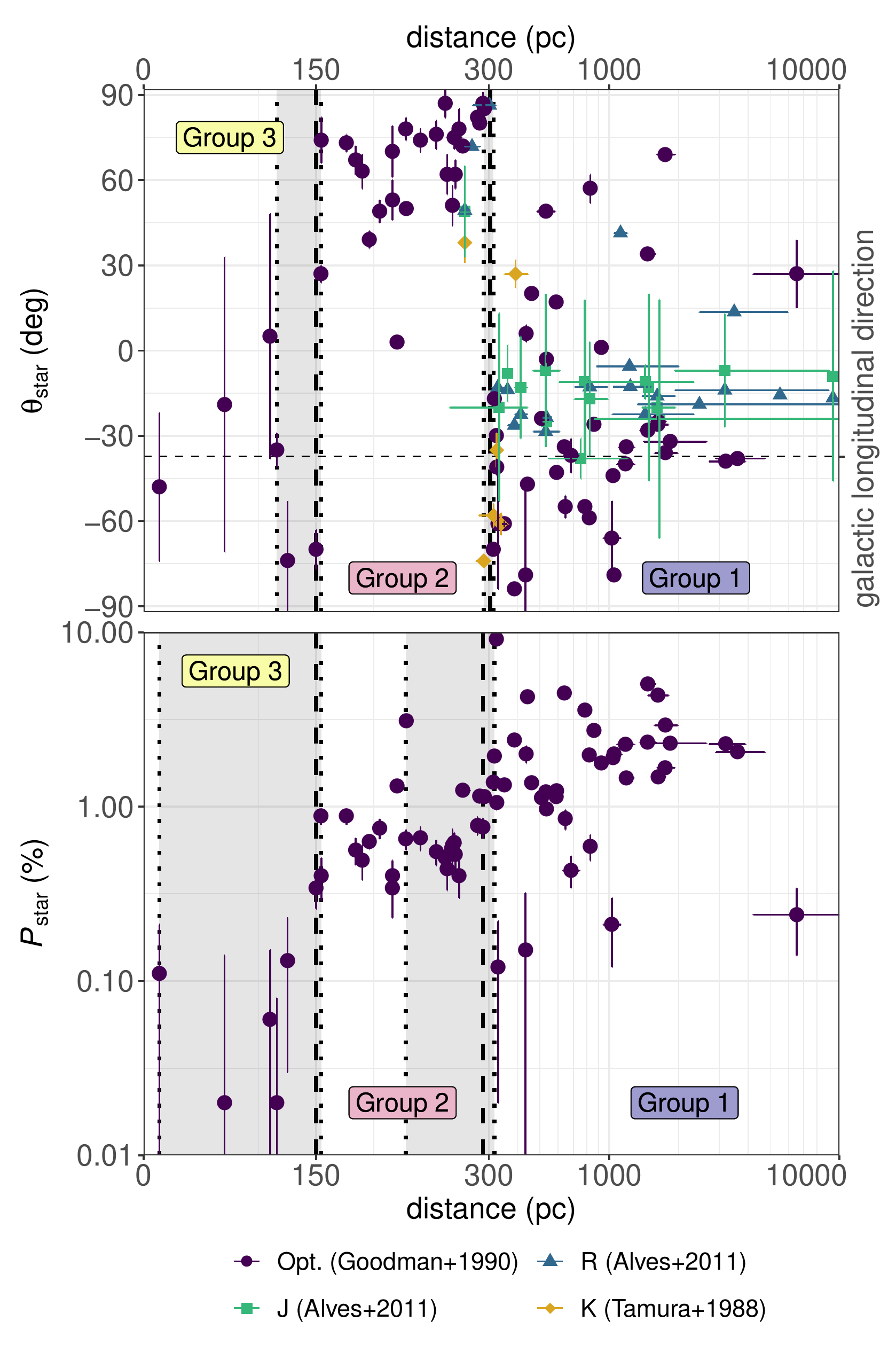}
\caption{The relationships between the stellar distances $d$ and $\theta_\mathrm{star}$ and $P_\mathrm{star}$.
Note that the horizontal axis is a linear scale below 300 pc, but a logarithmic scale above 300 pc.
Purple points are the optical polarimetry by \citet{1990ApJ...359..363G}.
Blue points are the \emph{R}-band polarimetry and green points are the \emph{J}-band polarimetry, respectively, by \citep{2011AJ....142...33A}.
Yellow points are \emph{K}-band polarimetry by \citet{1988MNRAS.231..445T}.
Three stellar groups based on their distances (Group 1, Group 2, and Group 3: see Section \ref{sec:breakpoints} and Table \ref{tab:ANG_P_Group}) are indicated.
Vertical dashed lines indicate the breakpoints in the distribution of the data points, and shaded area between dotted lines indicate their 95\% confidence intervals.
See Section \ref{sec:breakpoints} for the detailed description of the breakpoints estimation.
The horizontal dashed line in the top panel shows the orientation parallel to the galactic plane.
}
\label{fig:AngPbreak}
\end{figure}

In Figure \ref{fig:AngPbreak}, we display the $\theta_\mathrm{star}$ and $P_\mathrm{star}$ dependences as a function of the estimated stellar distances.
As seen in the figure, there is a clear jump in both distributions at a distance of about 300 pc, which is the distance to the Perseus molecular cloud.
In addition to that, another jump is noticeable at a distance of about 150 pc.
Hereafter, we call the polarimetry data whose stellar distances $d > 300$ pc as Group 1, those with $150 < d < 300$ pc as Group 2, and those with $d < 150$ pc as Group 3, respectively.
We summarize $\theta_\mathrm{star}$ and $P_\mathrm{star}$ values of each group in Table \ref{tab:ANG_P_Group}.
\begin{table}[t]
\begin{center}
\caption{$\theta_\mathrm{star}$ and $P_\mathrm{star}$ Values of Each Distance Group obtained from Optical Polarimetry Data \citep{1990ApJ...359..363G}}
\label{tab:ANG_P_Group}
\begin{tabular}{lcccc}
\hline
\hline
& Distance & $N_\mathrm{star}$$^\mathrm{a}$ & $\theta_\mathrm{star}$$^\mathrm{b}$ & $P_\mathrm{star}$\\
& (pc) & & (deg) & (\%)\\
\hline
Group 1 & $> 300$ & 39 & $-37.6 \pm 35.2$ & $2.0 \pm 1.7$ \\
Group 2 & 150 -- 300 & 25 & $+66.8 \pm 19.1$ & $0.8 \pm 0.6$ \\
Group 3 & $< 150$ & 6 & $-41.8 \pm 32.0$ & $0.1 \pm 0.1$ \\
\hline
\multicolumn{5}{p{0.44\textwidth}}{$^\mathrm{a}$The number of stars.}\\
\multicolumn{5}{p{0.44\textwidth}}{$^\mathrm{b}$$\theta$'s mean and standard deviation values are circular means and circular standard deviations, which take into account the $\theta$'s $180^\circ$ degeneracy, throughout this paper.
The definitions of the circular mean and the circular standard deviation are given by \citet{2020ApJ...899...28D}.}
\end{tabular}
\end{center}
\end{table}

Figure \ref{fig:AngPbreak} shows that $\theta_\mathrm{star}$ and $P_\mathrm{star}$ are both consistent within their own distance group.
That is, the stars in Group 1 and the stars in Group 2 are tracing the polarization from ISM clouds located at distances of 300 pc and 150 pc, respectively.
In particular, the consistency of $P_\mathrm{star}$ indicates that the ISM between and behind the two clouds does not significantly contribute to stellar polarization.
Group 3 can be thought as foreground stars with little interstellar extinction.
We note that these stars have very high uncertainties in their polarization measurements due to having very low polarization fractions.

To quantitatively estimate the distance of the two ISM clouds that cause polarization, we perform a breakpoint analysis on $\theta_\mathrm{star}$ and $P_\mathrm{star}$ distributions as a function of $d$ shown in Figure \ref{fig:AngPbreak}.
We assume that $\theta_\mathrm{star}$ and $P_\mathrm{star}$ are constant as a function of $d$, which corresponds to the assumption that the observed polarization is caused by 2D sheet(s) of ISM at specific distance(s).
In addition, we assume the $\theta_\mathrm{star}$ and $P_\mathrm{star}$ distributions have a certain number of step-wise changes (i.e., breakpoints), which correspond to the positions of the 2D sheets.
We perform least-squares fits to the data and make most likelihood estimations (MLE) of the positions of breakpoints.
We then repeat the fit with different number of breakpoints, and compare the goodness-of-fit values based on the Bayesian information criterion, to get the most likely numbers of breakpoints and their positions.

We perform the breakpoint analysis for each distance dependence of $\theta_\mathrm{star}$ and $P_\mathrm{star}$ shown in Figure \ref{fig:AngPbreak} by using the R library `strucchange' \citep{strucchange,breakpoints}.
For $\theta_\mathrm{star}$ distribution analysis, we combine optical and NIR polarimetry data as shown in Figure \ref{fig:AngPbreak} and use the shortest waveband data if multiple waveband data are available for a single star.
On the other hand, we use only optical polarimetry data to analyze $P_\mathrm{star}$ distribution because different wavelengths give different polarization fractions.
We used $P_\mathrm{star}$'s logarithm values in our analysis to detect breakpoints of widely different $P_\mathrm{star}$ values (see Figure \ref{fig:AngPbreak}) with comparable sensitivity to each other.

The estimated distances of the breakpoints are shown in Figure \ref{fig:AngPbreak} and Table \ref{tab:ANG_P_breaks}.
The breakpoint analysis shows that there are two breakpoints in both $\theta_\mathrm{star}$ and $P_\mathrm{star}$ distributions.
The estimated distances are $150^{+4}_{-34}$ pc and $303^{+12}_{-7}$ pc for $\theta_\mathrm{star}$, and $150^{+4}_{-136}$ pc and $295^{+22}_{-67}$ pc for $P_\mathrm{star}$, respectively.
Here we show the MLE positions and their 95\% confidence intervals.
Although the breakpoints in $P_\mathrm{star}$ have larger errors comparing to that in $\theta_\mathrm{star}$ due to smaller jumps in $P_\mathrm{star}$ values, the estimated distances are fully compatible between $\theta_\mathrm{star}$ and $P_\mathrm{star}$.
\begin{table}[t]
\begin{center}
\caption{Estimated Cloud Distances as $\theta_\mathrm{star}$ and $P_\mathrm{star}$ Breakpoints}
\label{tab:ANG_P_breaks}
\begin{tabular}{lll}
\hline
\hline
& \multicolumn{2}{c}{Cloud Distance}\\
& \multicolumn{2}{c}{(pc)}\\
\hline
$\theta_\mathrm{star}$ & $150^{+4}_{-34}$ & $303^{+12}_{-7}$ \\
$P_\mathrm{star}$ & $150^{+4}_{-136}$ & $295^{+22}_{-67}$ \\
\hline
\end{tabular}
\end{center}
\end{table}

The distance of the breakpoint at 300 pc is consistent with the distance to the Perseus molecular cloud \citep{2018ApJ...865...73O, 2018ApJ...869...83Z,2021A&A...645A..55P}.
Thus, we conclude that the polarization of Group 1 traces the \emph{B}-field of the Perseus molecular cloud.
The distance of another breakpoint at 150 pc is thought to indicate the distance to the foreground ISM cloud that causes the polarization of Group 2.
This 150 pc foreground ISM also contribute to depolarize Group 1.
We will discuss this depolarization effect in Section \ref{sec:qu_separation}.
Group 3 are the foreground stars that show no clear indication of interstellar extinction.

\subsection{Estimation of Cloud Distances based on Photometry and Gaia Distances}
\label{sec:Gaia_Phot}

In the previous section (Section \ref{sec:breakpoints}), we estimated the distances of the breakpoints found in polarimetric data.
The accuracy of estimation is limited to a few tens of parsecs because of small number of data points.
In this section, we perform a breakpoint analysis using all the available stellar photometric data in the Perseus molecular cloud direction to make a more accurate estimation of the distances to the breakpoints, including the foreground ISM.
We then compare our result with other existing estimates of cloud distances in this direction to deomonstrate the reliability of our method for measuring dust clouds' distance.

\begin{figure}[tp]
\centering
\includegraphics[width=\linewidth]{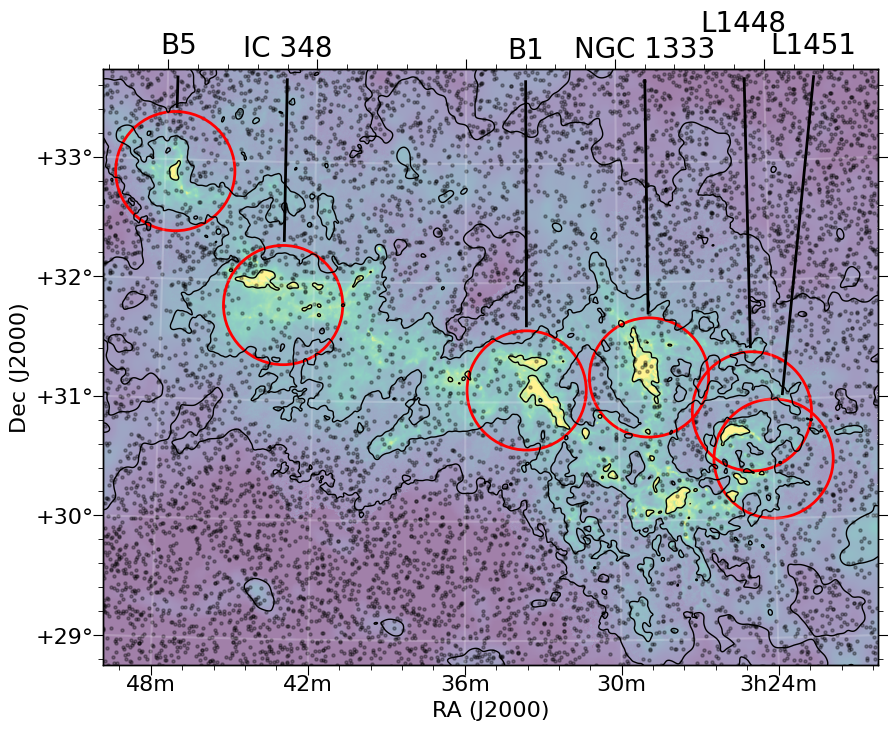}
\caption{
Distribution of Gaia photometric data in the direction of the Perseus molecular cloud.
We plot the stars whose distances $d \leq 1$ kpc and $\mathrm{RUWE} \leq 1.4$ (see Section \ref{sec:DR2} for the definition of RUWE).
The background color scale and contours are the hydrogen column density estimated by Herschel and Planck observational data \citep[][see Figure \ref{fig:Goodman}]{2016A&A...587A.106Z}.
Contour levels are $N_\mathrm{H} = 0.5,~1,~\mathrm{and}~5 \times 10^{22} ~\left( \mathrm{cm}^{-2} \right)$.
Red circles (diameter: $1^\circ$) are the regions where we estimate the breakpoint distances and compare them with the CO molecular cloud distances estimated by \citet{2018ApJ...869...83Z}.
Names of the CO molecular clouds are shown above the figure.
See Section \ref{sec:Gaia_Phot} and Figure \ref{fig:COMPLETE_distance} for the details of the comparison.}
\label{fig:PerseusGaiadist}
\end{figure}

We use \emph{G}-band extinction ($A_G$) values catalogued in Gaia DR2 with $d \leq 1$ kpc and $\mathrm{RUWE} \leq 1.4$, and fit these values as a function of distances obtained from Gaia parallax.
The spatial distribution of the stellar data used in this analysis is shown in Figure \ref{fig:PerseusGaiadist}.
We note the paucity of stars in the direction of the dense molecular cloud.
This is due to the larger extinction in the visible \emph{G}-band compared with the NIR bands.

\citet{2018ApJ...869...83Z} combined Gaia distance and NIR photometry to obtain the distance of each velocity component of the $^{12}\mathrm{CO}~(1-0)$ line emission for each cloud core in the Perseus molecular cloud.
We perform breakpoint analyses for similar spatial directions and compare the result with their estimation.
Since the exact spatial region analyzed in \citet{2018ApJ...869...83Z} is not indicated in the literature, we estimate the breakpoints at $1^\circ$ diameter regions centered at each molecular cloud core.
The regions are shown as red circles in Figure \ref{fig:PerseusGaiadist}.

\begin{figure}[tp]
\centering
\includegraphics[width=\linewidth]{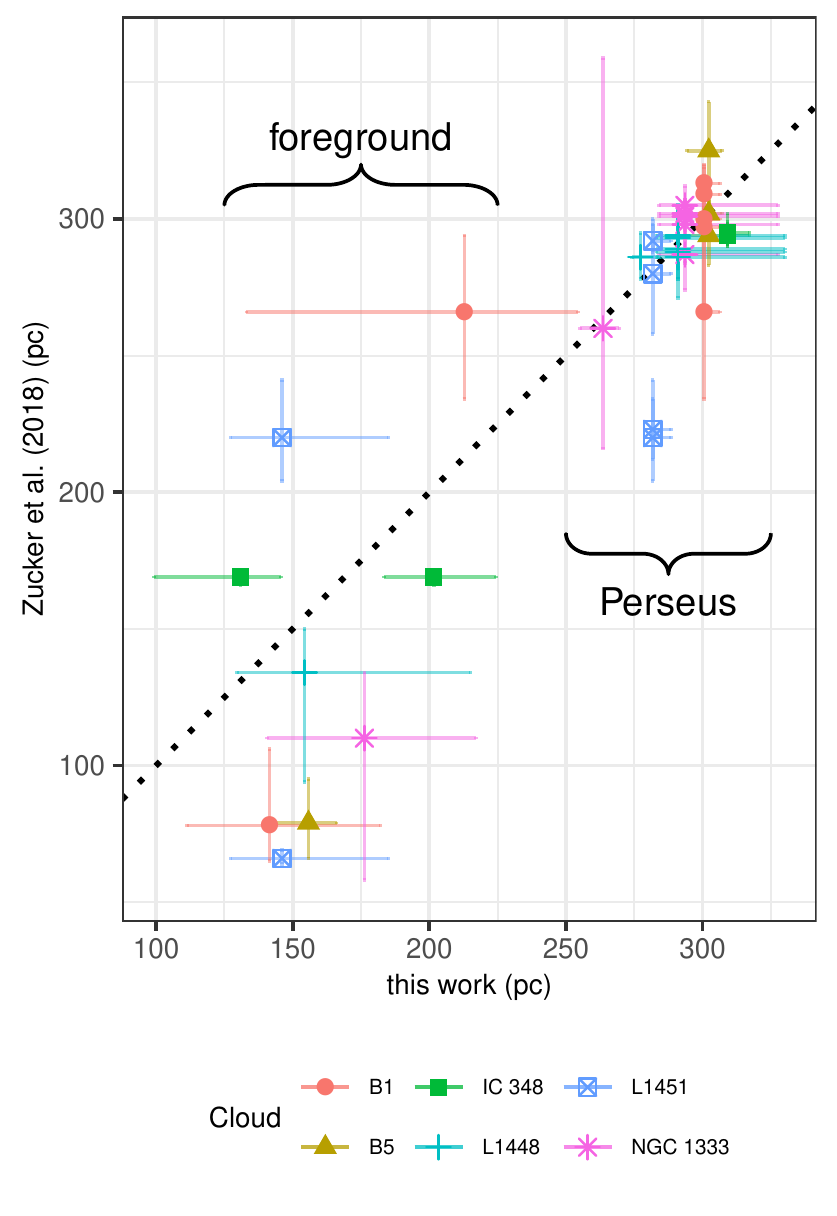}
\caption{
Estimated distances of major star-forming regions across the Perseus molecular cloud.
Results of our breakpoint analysis (`this work' in the figure) are compared with the CO distances by \citet[][also see Table \ref{tab:COMPLETE_distance}]{2018ApJ...869...83Z}.
Our breakpoint analysis identifies multiple clouds in each region (Table \ref{tab:COMPLETE_distance}, 4th -- 6th columns).
On the other hand, \citet{2018ApJ...869...83Z} identifies distances for each velocity component of the $^{12}\mathrm{CO}~(1-0)$ emission line (Table \ref{tab:COMPLETE_distance}, 7th -- 13th columns).
Therefore, the clouds identified by this work and \citet{2018ApJ...869...83Z} do not necessarily have a one-to-one correspondence.
For simplicity, we show the closest pair of clouds by \citet{2018ApJ...869...83Z} in the same region for each cloud identified by the breakpoint analysis.
Similarly, for each cloud by \citet{2018ApJ...869...83Z}, the closest cloud in the same region by the breakpoint analysis is shown as a pair.
Therefore, note that the same cloud may be paired with two different clouds and shown twice in the figure.
}
\label{fig:COMPLETE_distance}
\end{figure}

\begin{table*}[t]
\begin{center}
\caption{Estimated Distances of Major Star-forming Regions across the Perseus Molecular Cloud}
\label{tab:COMPLETE_distance}
\begin{tabular}{lrr|rrr|rrrrrrr}
\hline
\hline
Cloud Name & R.A. & Decl. & \multicolumn{3}{c|}{This Work} & \multicolumn{7}{c}{\citet{2018ApJ...869...83Z}$^\mathrm{a}$}\\
& (deg) & (deg) & \multicolumn{3}{c|}{(pc)} & \multicolumn{7}{c}{(pc)} \\
\hline
L1451 & 51.0 & 30.5 & $146^{+39}_{-19}$ & ~ & $282^{+7~}_{-0~}$ & $66^{+3~}_{-2~}$ & $220^{+21}_{-16}$ & $223^{+11}_{-11}$ & $280^{+20}_{-22}$ & $292^{+6~}_{-4~}$ & & \\
L1448 & 51.2 & 30.9 & $154^{+61}_{-25}$ & $277^{+7~}_{-3~}$ & $290^{+7~}_{-3~}$ & $134^{+16}_{-40}$ & $286^{+9~}_{-8~}$ & $288^{+15}_{-17}$ & $289^{+10}_{-11}$ & $293^{+10}_{-11}$ & $294^{+11}_{-12}$ & \\
NGC 1333 & 52.2 & 31.2 & $176^{+41}_{-36}$ & $264^{+6~}_{-8~}$ & $294^{+34}_{-9~}$ & $110^{+24}_{-52}$ & $260^{+99}_{-44}$ & $287^{+20}_{-13}$ & $298^{+9~}_{-14}$ & $301^{+4~}_{-3~}$ & $302^{+5~}_{-6~}$ & $305^{+7}_{-7}$ \\
B1 & 53.4 & 31.1 & $142^{+41}_{-30}$ & $213^{+41}_{-30}$ & $300^{+6~}_{-1~}$ & $78^{+28}_{-13}$ & $266^{+28}_{-32}$ & $297^{+23}_{-33}$ & $300^{+4~}_{-4~}$ & $309^{+5~}_{-8~}$ & $313^{+6~}_{-5~}$ & \\
IC 348 & 55.8 & 31.8 & $131^{+15}_{-32}$ & $202^{+15}_{-32}$ & $309^{+8~}_{-2~}$ & $169^{+2}_{-3}$ & $294^{+5~}_{-4~}$ & $295^{+7~}_{-5~}$ & $295^{+5~}_{-4~}$ & $295^{+7~}_{-4~}$ & & \\
B5 & 56.9 & 32.9 & $156^{+10}_{-13}$ & ~ & $302^{+5~}_{-8~}$ & $79^{+16}_{-13}$ & $294^{+13}_{-11}$ & $302^{+8~}_{-9~}$ & $325^{+18}_{-17}$ & & & \\
\hline
\multicolumn{13}{p{0.85\textwidth}}{\bf Notes.}\\
\multicolumn{13}{p{0.85\textwidth}}{$^\mathrm{a}$Estimated distances to each velocity component of the CO molecular line emission for each cloud \citep{2018ApJ...869...83Z} is shown. The leftmost (closest) component of each cloud are foreground components.}\\
\end{tabular}
\end{center}
\end{table*}

We show the comparison of the results of our breakpoint analysis with the estimation by \citet{2018ApJ...869...83Z} in Figure \ref{fig:COMPLETE_distance} and Table \ref{tab:COMPLETE_distance}.
Our results are in reasonable agreement with those of \citet{2018ApJ...869...83Z} for each component of the Perseus cloud at a distance of about 300 pc.
The estimated distances show a gradual increase from west to east as pointed out by \citet[][also see \citealp{2019ApJ...877...88C}]{2018ApJ...869...83Z}.
Therefore, we can conclude that our \emph{G}-band breakpoint analysis can measure the distance to the cloud with sufficient accuracy.

\begin{figure}[tp]
\centering
\includegraphics[width=\linewidth]{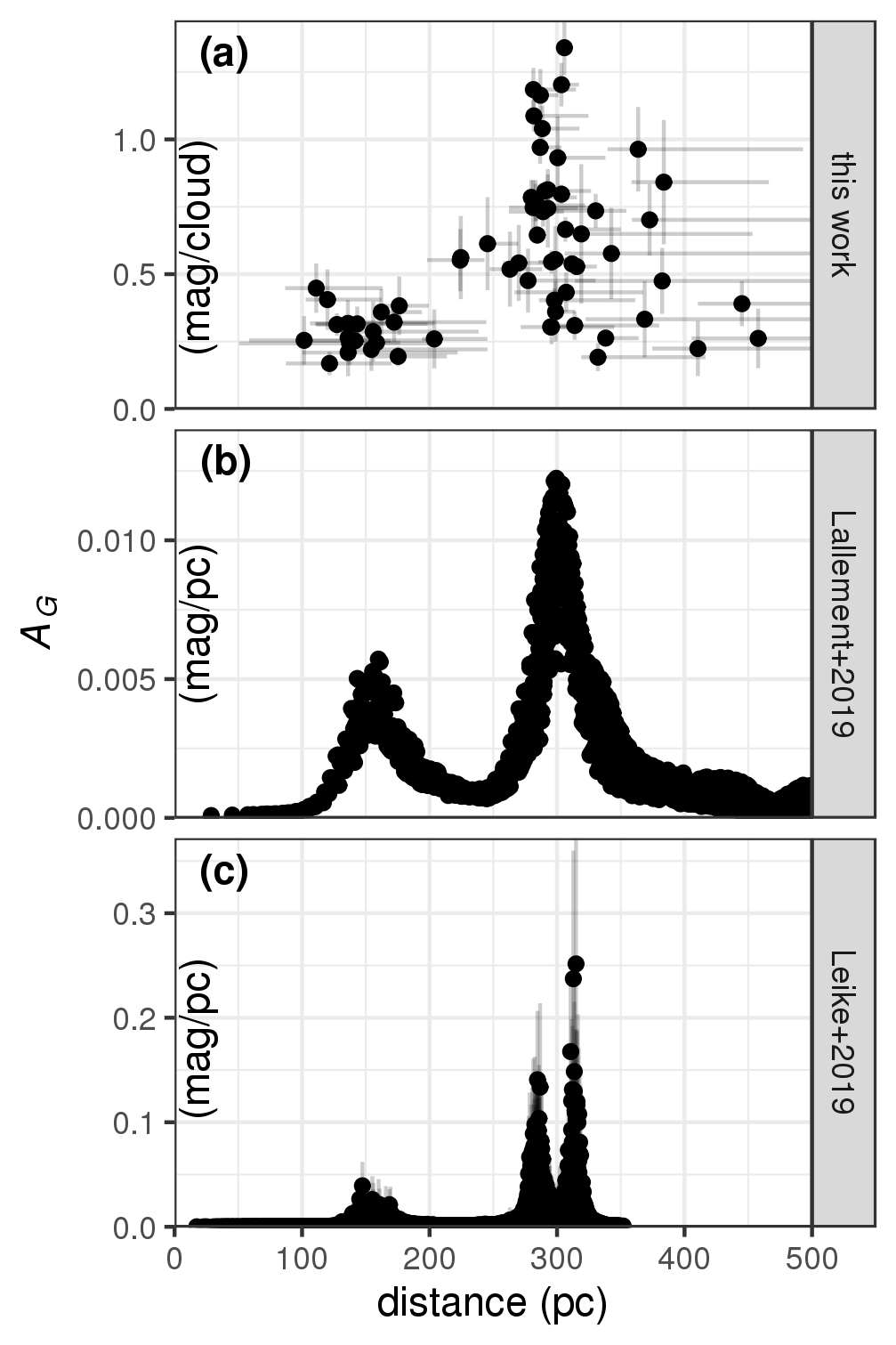}
\caption{
Comparison of the distributions of the breakpoints distances in the Perseus molecular cloud direction ($157^\circ < l < 161^\circ,~-23^\circ < b < -16^\circ$) with existing 3D dust maps as a function of the distance.
Vertical axes are for the \emph{G}-band extinction ($A_G$) per each detected cloudlet (i.e., breakpoint) for panel (a), and $A_G$ per pc for panels (b) and (c).
}
\label{fig:cloud_distance_profile}
\end{figure}

On the other hand, the position of the foreground component is less consistent.
Therefore, we further check our estimated foreground distance's consistency, comparing that with 3D dust maps based on Gaia distance data by \citet{2019A&A...625A.135L} and \citet{2019A&A...631A..32L}.
We show a comparison between these results and ours for the Perseus molecular cloud direction in Figure \ref{fig:cloud_distance_profile}.

Figure \ref{fig:cloud_distance_profile}a shows the results of our breakpoint analysis.
The spatial resolution of this analysis is set as $55\arcmin$.
The estimated cloud distances are concentrated at 150 pc and 300 pc (Figure \ref{fig:cloud_distance_profile}a).

Figures \ref{fig:cloud_distance_profile}b and c show the results by \citet{2019A&A...625A.135L} and \citet{2019A&A...631A..32L}, respectively.
Both two dust maps successfully detect the foreground cloud in addition to the Perseus cloud.
\citet[][Figure \ref{fig:cloud_distance_profile}b]{2019A&A...625A.135L} combine 2MASS photometric data with Gaia DR2.
Their map has limited LOS resolution of 50 pc and thus the estimated extinction for the Perseus cloud is smoothed out, resulting in relatively lower extinction per unit length value ($\sim 0.01$ mag/pc) compared to \citet[][Figure \ref{fig:cloud_distance_profile}c]{2019A&A...631A..32L}.
\citet{2019A&A...625A.135L} indicated plans to improve the LOS resolution as the Gaia data is updated.

\citet[][Figure \ref{fig:cloud_distance_profile}c, also see \citealp{2020A&A...639A.138L}]{2019A&A...631A..32L} used the Gaia DR2 $A_G$ values as photometric data to create a 3D dust map of the solar vicinity ($d \lesssim 300$ pc).
They achieved higher resolution in LOS comparing to the map by \citet{2019A&A...625A.135L} that used NIR photometric data (Figure \ref{fig:cloud_distance_profile}b).
NIR photometry is effective for tracing dense molecular clouds' interior but has limited sensitivity for tracing faint clouds with a high spatial resolution.
On the other hand, the disadvantage of tracing the shape of the dust cloud using only photometry of visible wavebands is also apparent; in their analysis shown in Figure \ref{fig:cloud_distance_profile}c, the 300 pc Perseus molecular cloud cannot be correctly detected due to saturation of the $A_G$ value and is split into two distances before and after the cloud.

Our results shown in Figure \ref{fig:cloud_distance_profile}a are based on the simple 2D sheet assumption and have limited spatial resolution of about $1^\circ$, but the use of the $A_G$ value achieves a good sensitivity to faint clouds as well as the correct determination of the distances of dense molecular clouds.

According to the discussions above, we judge that our method is a simple and effective one for measuring dust clouds.
Using \emph{G}-band photometry, we can obtain the distance to both faint and dense clouds stably and accurately.
Our results described in Section \ref{sec:breakpoints} (Figure \ref{fig:AngPbreak}) are thus consistent that the stellar polarization in the Perseus molecular cloud's direction originates in two isolated clouds at 150 pc and 300 pc, which are both detected by \citet{2019A&A...625A.135L} and \citet{2019A&A...631A..32L}.

\subsection{Spatial Distribution of Individual Clouds}
\label{sec: Taurus foreground}

\begin{figure}[tp]
\centering
\includegraphics[width=\linewidth]{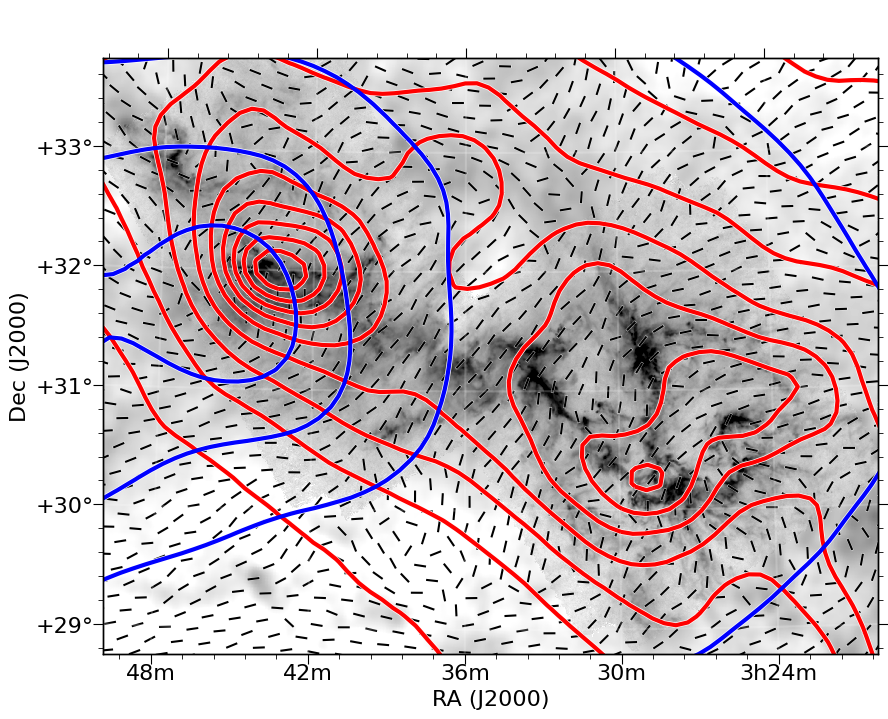}
\includegraphics[width=1.05\linewidth]{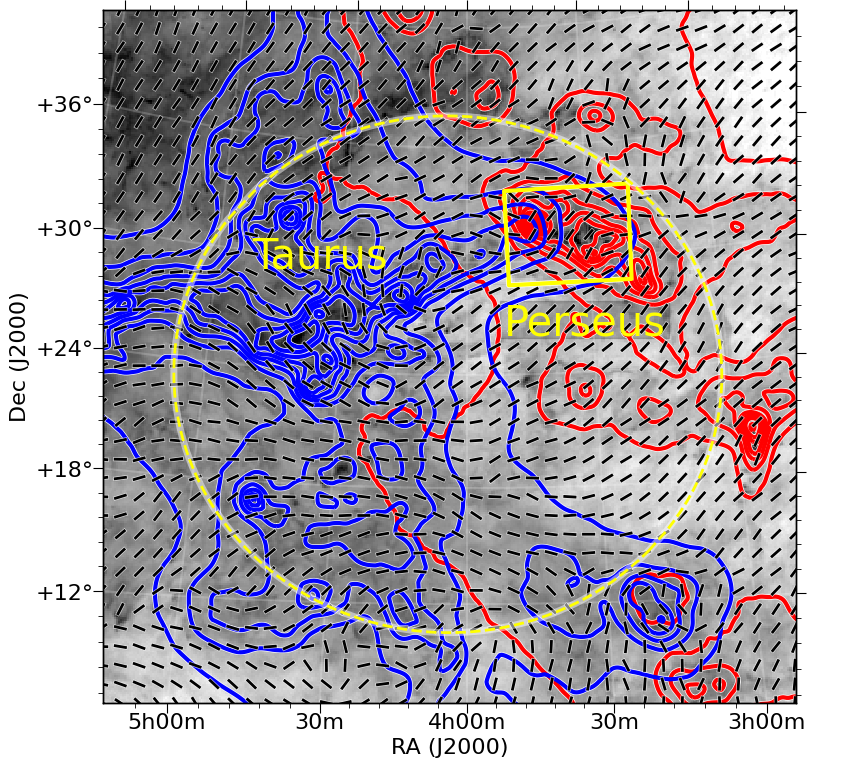}
\caption{
The spatial distribution of the dust clouds at 150 pc and 300 pc estimated by \citet{2019A&A...631A..32L}.
The blue contours are for the dust cloud integrated between 100 pc and 200 pc (i.e., the foreground cloud component), and the red contours are for the dust cloud integrated between 250 pc and 350 pc (i.e., the Perseus cloud itself).
The contour levels are [0.1, 0.3, 0.5, \ldots] mag in $A_G$.
{\bf Top panel}: the distribution in the Perseus region.
The display area of the figure is the same as that of Figures \ref{fig:Goodman} and \ref{fig:PerseusGaiadist}.
The background gray scale is the hydrogen column density ($N_\mathrm{H}$), and the black line segments are the \emph{B}-field orientation observed by Planck ($\theta_\mathrm{Planck}$; see Figure \ref{fig:Goodman}).
The spatial resolution of $\theta_\mathrm{Planck}$ is set as $10'$ in the top panel.
{\bf Bottom panel}: the distribution in the Taurus-Perseus region.
The yellow rectangle indicates the area shown in the top panel.
The yellow dashed circle shows the dust cavity outline identified in the 3D spatial distribution of dust clouds.
See Section \ref{sec:PerOB2} for more details on dust cavity identification.
The background gray scale is the Planck-observed 353 GHz continuum intensity.
The black line segments are $\theta_\mathrm{Planck}$, whose spatial resolution is set as $1^\circ$ in the bottom panel.
}
\label{fig:TAP40}
\end{figure}

Figure \ref{fig:TAP40} shows the spatial distribution of the dust cloud at 150 pc and 300 pc by \citet{2019A&A...631A..32L} as contours.
To avoid the effect of LOS splitting of the dense molecular cloud, we show the spatial distribution of the integrated $A_G$ magnitude in the 100--200 pc range for the 150 pc component and in the 250--350 pc range for the 300 pc component, respectively.
In addition to the Perseus cloud (the top panel of Figure \ref{fig:TAP40}), we also show the Taurus-Perseus molecular cloud complex in the bottom panel of Figure \ref{fig:TAP40}.
As is evident in Figure \ref{fig:TAP40}, the 150 pc foreground cloud lying in front of the Perseus molecular cloud corresponds to the outer edge of the Taurus molecular cloud at a distance of $\sim 140$ pc \citep{2019ApJ...885...19Y, 2019ApJ...879..125Z, 2020A&A...638A..85R}, as was proposed by \citet{1987ApJS...63..645U} and \citet{1990Ap&SS.166..315C}.
The estimated extinction of the foreground cloud based on \citet{2019A&A...631A..32L} is $A_G^\mathrm{150pc} = 0.1$ -- 0.9 mag. 

We overlay the \emph{B}-field orientation measured by Planck ($\theta_\mathrm{Planck}$) in Figure \ref{fig:TAP40} as black line segments.
$\theta_\mathrm{Planck}$ toward the Perseus cloud is generally aligned to the northwest-southeast direction.
The circular mean and the circular standard deviation are $\theta_\mathrm{Planck} = -56^\circ \pm 25^\circ$ if we estimate them  in the region where the 300 pc dust cloud component $A_G^\mathrm{300pc} > 0.6$ mag.
This $\theta_\mathrm{Planck}$ is consistent with the optical $\theta_\mathrm{star}$ of Group 1 ($-37.6^\circ \pm 35.2^\circ$).
Outside the Perseus molecular cloud but on the foreground component, we note a significant change in the \emph{B}-field orientation.
If we estimate $\theta_\mathrm{Planck}$ on the foreground cloud whose $A_G^\mathrm{150pc} > 0.2$ mag, $A_G^\mathrm{300pc} < 0.6$ mag, and R.A. $< 4^\mathrm{h}00^\mathrm{m}$, the position angle is $\theta_\mathrm{Planck} = -89^\circ \pm 19^\circ$, which is considerably different from $\theta_\mathrm{Planck}$ on the Perseus molecular cloud and is consistent with the $\theta_\mathrm{star}$ of Group 2 ($+66.8^\circ \pm 19.1^\circ$) within the range of errors.

As a result, we conclude that there are two \emph{B}-field components observed in the Perseus molecular cloud's direction: one is the Perseus molecular cloud's \emph{B}-field, and the other is that of a tenuous cloud at the outer edge of the Taurus molecular cloud with $A_G < 1$ mag.

\subsection{Polarization of the two clouds}
\label{sec:qu_separation}

\begin{figure}[tp]
\centering
\includegraphics[width=\linewidth]{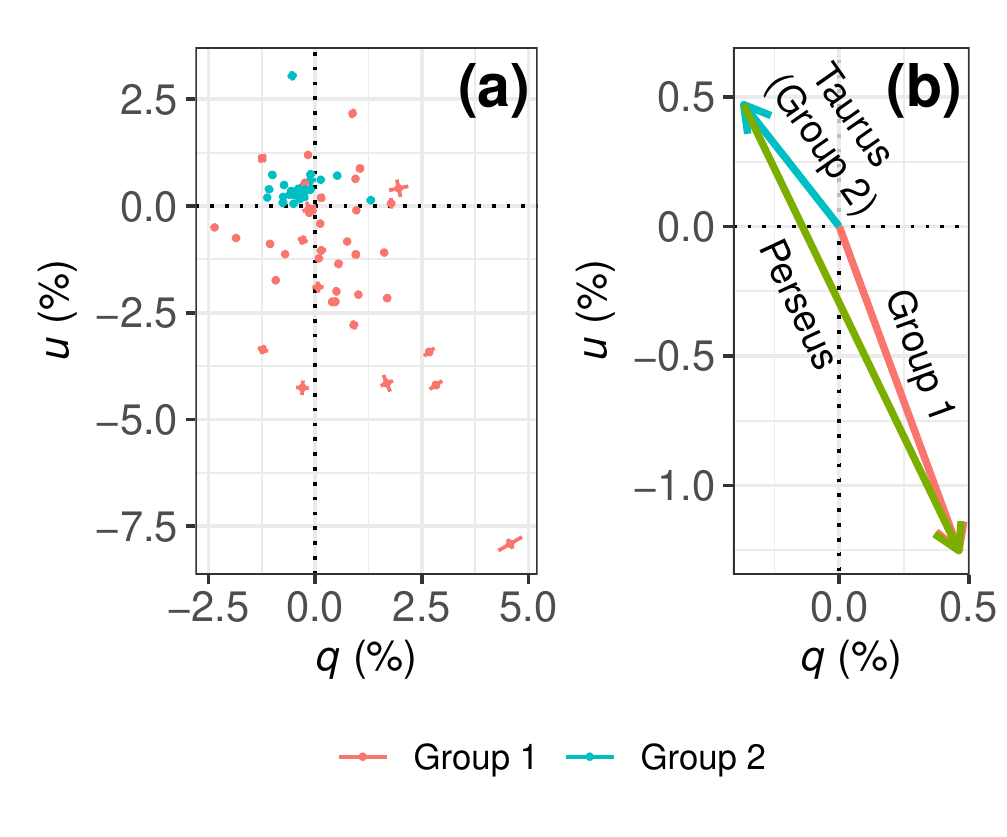}
\caption{
{\bf Left panel (a)}: the distribution of the observed relative Stokes $u$ and $q$ parameters, estimated from the observed $P_\mathrm{star}$ and $\theta_\mathrm{star}$ of Group 1 ($d > 300$ pc) and Group 2 ($150 < d < 300$ pc) stars.
The radial error bars show the observed error in $P_\mathrm{star}$, and the circumferential error bars show the observed error in $\theta_\mathrm{star}$.
The error bars may be hidden behind the marks.
{\bf Right panel (b)}: decomposition of the Group 1 $q$-$u$ vector into that of the Taurus cloud and the Perseus cloud.
The relationship between the average values of Taurus and Perseus clouds is shown.
}
\label{fig:QU_separation}
\end{figure}

The two-component \emph{B}-field of Perseus and Taurus clouds, described in the previous section, is traced by Group 1 ($d > 300$ pc) and Group 2 ($150 < d < 300$ pc) stellar polarization data.
Group 2 traces only the Taurus cloud, while Group 1 sees through both Perseus and Taurus clouds.
Here we estimate the Perseus molecular cloud's \emph{B}-field by removing the Taurus contribution from Group 1.

The observed polarization fraction is $\ll 10$ \% (Figure \ref{fig:AngPbreak} and Table \ref{tab:ANG_P_Group}).
In such a low polarization condition, the relative Stokes parameters $q$ (= Q/I) and $u$ (= U/I) in the polarized flux can be approximated as additive (e.g., \citealp{2019ApJ...872...56P} and the references therein).
In other words, we can separate the contribution to the polarization from individual clouds as follows.
\begin{eqnarray}
q_\mathrm{\,Group1} &=& q_\mathrm{\,Taurus} + q_\mathrm{\,Perseus}, \nonumber\\
q_\mathrm{\,Group2} &=& q_\mathrm{\,Taurus}, \nonumber
\end{eqnarray}
where $q_\mathrm{\,Group1}$ and $q_\mathrm{\,Group2}$ are the relative Stokes $q$ parameter observed for Group 1 and Group 2 stars, and $q_\mathrm{\,Taurus}$ and $q_\mathrm{\,Perseus}$ are the relative Stokes $q$ parameter originated in Taurus and Perseus clouds, respectively.
These formulas also hold for the relative Stokes $u$ parameter.
We estimate $q_\mathrm{\,star}$ and $u_\mathrm{\,star}$ values of Group 1 and Group 2 from the observed $\theta_\mathrm{star}$ and $P_\mathrm{star}$ as follows.
\begin{eqnarray}
q_\mathrm{\,star} &=& P_\mathrm{star} \times \cos\left(2 \cdot \theta_\mathrm{star}\right),\nonumber \\
u_\mathrm{\,star} &=& P_\mathrm{star} \times \sin\left(2 \cdot \theta_\mathrm{star}\right).\nonumber
\end{eqnarray}
The estimated $q_\mathrm{\,star}$ and $u_\mathrm{\,star}$ are shown in Figure \ref{fig:QU_separation}a.

The $q$ and $u$ values of Group 1 show a significant scatter, reflecting the local variation in $\theta_\mathrm{star}$ and $P_\mathrm{star}$ of Group 1.
On the other hand, the data of Group 2, which represents the $q$ and $u$ values of the Taurus cloud, show a small variation.
This is because $\theta_\mathrm{star}$ and $P_\mathrm{star}$ of Group 2 have a small spatial variation, and the value of $P_\mathrm{star}$ is also small (see Table \ref{tab:ANG_P_Group}).

Since $q$ and $u$ are additive, the observed values of $q$ and $u$ in Group 1 can be expressed as a vector sum of the contributions from Perseus and Taurus clouds on the $q$-$u$ plane.
This relationship is shown in Figure \ref{fig:QU_separation}b.
Here we show the relationship between the averaged $q$ and $u$ values of Taurus and Perseus.

The estimated $q$-$u$ vectors of Perseus and Taurus are nearly opposite to each other.
That is, the estimated orientations of the POS \emph{B}-field of these two clouds are nearly perpendicular to each other ($\theta_\mathrm{Perseus} \perp \theta_\mathrm{Taurus}$).
Furthermore, the Perseus contribution to the Group 1 $q$-$u$ vector is dominant compared to that of Taurus.
This may reflect the fact that the column density of the Perseus main cloud is dominant compared to that of the foreground Taurus cloud's outskirt (see Figure \ref{fig:cloud_distance_profile}).
As a result, the Taurus contribution to the Group 1 polarization does not significantly change the Perseus' q-u vector direction (Figure \ref{fig:QU_separation}b).
Thus, we conclude the followings.
\begin{eqnarray}
\theta_\mathrm{Perseus} &\simeq& \theta_\mathrm{Group1}, \nonumber\\ \theta_\mathrm{Taurus} &=& \theta_\mathrm{Group2}. \nonumber
\end{eqnarray}
On the other hand, we estimate that $P_\mathrm{Perseus}$ is slightly larger than $P_\mathrm{Group1}$ due to the depolarization by the foreground Taurus cloud.
We estimate $\theta_\mathrm{Perseus}$ and $P_\mathrm{Taurus}$ by subtracting the averaged Group 2 $q$-$u$ vector from the observed Group 1 $q$-$u$ vectors, as shown in Figure \ref{fig:QU_separation}b.
The results are shown in Table \ref{tab:p_efficiency}.

\begin{table}[t]
\begin{center}
\caption{Polarization properties of individual clouds}
\label{tab:p_efficiency}
\begin{tabular}{lcccc}
\hline
\hline
Cloud & $\theta_\mathrm{star}$ & $P_\mathrm{star}$ & $A_G$ & $P_\mathrm{star}/A_G$\\
& (deg) & (\%) & (mag) & (\%/mag) \\
\hline
Taurus & $+66.8 \pm 19.1$ & $0.8 \pm 0.6$ & $0.32 \pm 0.21$ & $1.5 \pm 0.3$\\
Perseus & $-30.0 \pm 25.2$ & $2.4 \pm 1.8$ & $1.61 \pm 0.58$ & $1.5 \pm 0.2$\\
\hline
\end{tabular}
\end{center}
\end{table}

\begin{figure}[tp]
\centering
\includegraphics[width=\linewidth]{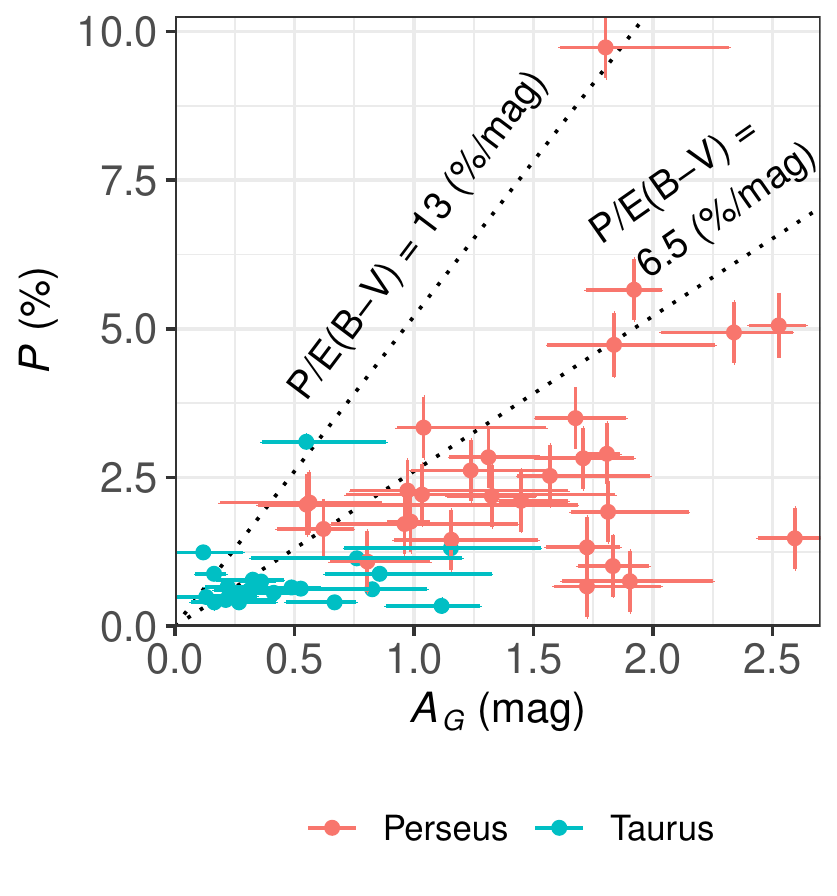}
\caption{
$P_\mathrm{star}$ of Perseus and Taurus clouds as a function of $A_G$.
See text for the estimation of $P_\mathrm{star}$ and $A_G$ of individual clouds.
The dashed lines show the observed maximum polarization efficiency ($P/E(B-V) = 13 \%$; \citealp{2019A&A...624L...8P,2020A&A...641A..12P}) and its 50 \% slope.
We assume $A_G/A_V = 0.789$ and $A_V/E(B-V) = 3.16$ \citep{2019ApJ...877..116W} to convert $E(B-V)$ to $A_G$.
}
\label{fig:pol_efficiency}
\end{figure}

Figure \ref{fig:pol_efficiency} shows $P_\mathrm{Perseus}$ and $P_\mathrm{Taurus}$ as a function of $A_G$ to investigate individual clouds' polarization efficiency ($P_\mathrm{star}/A_G$).
We assume that the Taurus cloud's $A_G$ values are equal to the Group 2 data.
For the Perseus cloud, we estimate $A_G$ by subtracting that of 150 pc cloud component estimated by \citet[][Figure \ref{fig:TAP40}]{2019A&A...631A..32L} at each position of the stars from the observed Group 1 $A_G$ values.
Note that the maximum observed $A_G$ value is 3 mag (also see Figure \ref{fig:offset_AG} and the discussion in Sections \ref{sec:PlanckCorrelation} and \ref{sec:small_scale}), indicating that the stellar extinction data are biased to regions in the cloud where the extinction is smaller ($A_G < 3$ mag).

In Figure \ref{fig:pol_efficiency}, we also show the observed maximum polarization efficiency taken from the literature ($P/E(B-V) = 13 \%$;  \citealp{2019A&A...624L...8P,2020A&A...641A..12P}).
The polarization efficiency of Taurus and Perseus corresponds to less than $\sim 50\%$ of the maximum efficiency.
We do not find a significant difference between the polarization efficiencies of Taurus and Perseus clouds.

\section{Discussion}
\label{sec:discussion}

\subsection{\emph{B}-field Orientation and the Per OB2 Bubble}
\label{sec:PerOB2}

In Section \ref{sec: Taurus foreground}, we identified that the 150 pc foreground cloud lying in front of the Perseus molecular cloud is the outer edge of the Taurus molecular cloud.
Both Taurus and Perseus molecular clouds are thought to be on a common large H\small{I} shell's outer edge \citep{1974IAUS...60..115S,2006A&A...451..539S,2019A&A...623A..16S}.
In Figure \ref{fig:cloud_distance_profile}, we find a low-extinction region between these two clouds.
In the following, we estimate the spatial extent of this low-extinction region.

\begin{figure*}[tp]
\centering
\includegraphics[width=\linewidth]{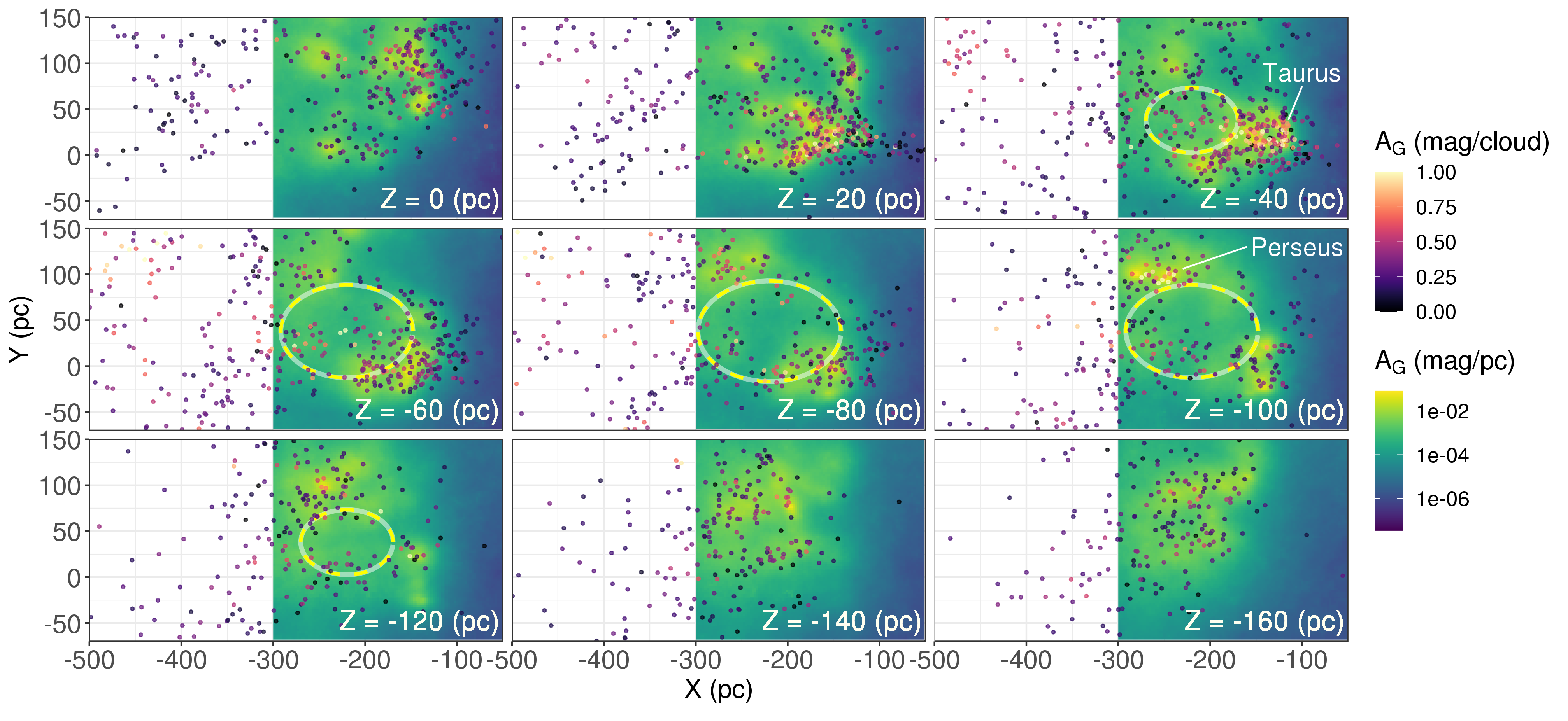}
\caption{
The 3D distribution of dust clouds in the Taurus-Perseus region.
Coordinates are the heliocentric galactic cartesian coordinates; 
the X-axis points to the galactic center, the Y-axis points to the direction of galactic rotation (galactic plane $l=90^\circ$), and the Z-axis points to the galactic north pole.
Points are the cloud positions determined by our breakpoint analysis, and the color scale is the distribution of dust clouds estimated by \citet[][available for the regions $\mathrm{X} \geq -300$ pc]{2019A&A...631A..32L}.
The distribution from parallel to the galactic plane at Z=0 pc to Z=-160 pc at every 20 pc is shown.
Yellow dashed ellipses show the cross-sectional profiles of the dust cavity (see Section \ref{sec:PerOB2}).
}
\label{fig:Perseus_chamber}
\end{figure*}

\begin{figure*}[tp]
\centering
\includegraphics[width=\linewidth]{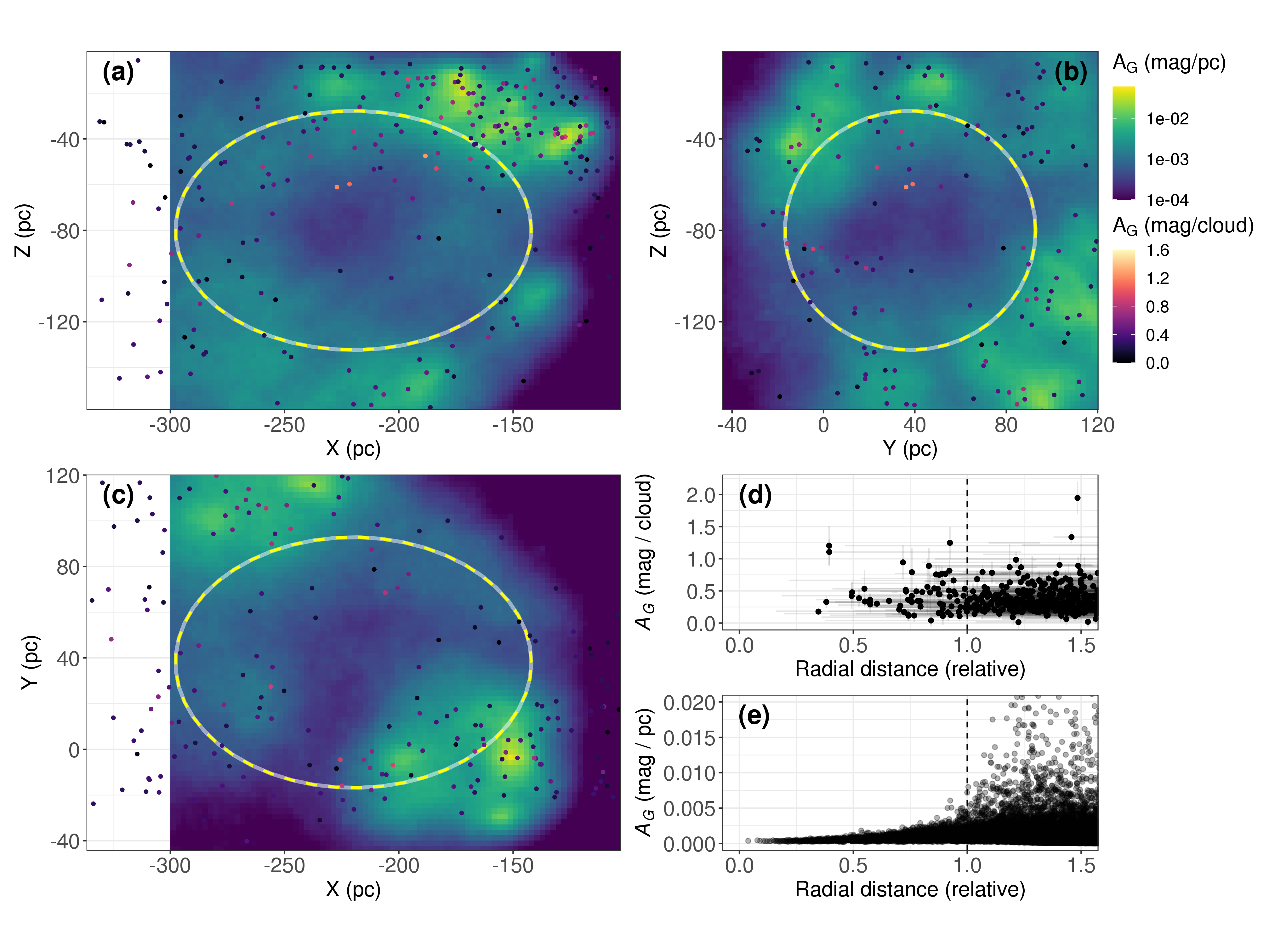}
\caption{
Cross-sectional views of the fitted ellipsoid to the dust shell are shown as dashed ellipses in panels (a): cross-sections in the direction of Y-axis, (b): same in X-axis, (c): same in Z-axis.
We show the cross sections in X, Y, and Z directions at the center of the fitted ellipsoid at $X = -220$ pc, $Y = 38$ pc, and $Z = -80$ pc.
The central position corresponds to $l = 170^\circ$, $b=-20^\circ$ in the galactic coordinates at the distance $d=240$ pc.
See section \ref{sec:PerOB2} for the description of the ellipsoid fit.
See Figure \ref{fig:Perseus_chamber} for the description of coordinates, color scales, and markers.
Dotted ellipses show the cross-sectional profiles of the dust cavity (see Section \ref{sec:PerOB2}).
The bottom right figures are radial profiles of the ISM, normalized to the radii of the fitted ellipsoid (the vertical dashed lines at radial distance = 1.0 correspond to the ellipsoid size in each direction).
Panel (d) is the distribution based on the breakpoint analysis, and panel (e) is that based on \citet{2019A&A...631A..32L}.
}
\label{fig:Perseus_chamber_section}
\end{figure*}

Figure \ref{fig:Perseus_chamber} shows the 3D distribution of the dust cloud in the region containing Taurus and Perseus.
We estimate the 3D shape of the dust shell by using the cloud positions determined by our breakpoint analysis.
We assume the shell's profile as a simple ellipsoid and perform a least-square fit of the shell to the cloud positions weighted by their extinction (mag/cloud).
The estimated size and position of the shell are shown in Figure \ref{fig:Perseus_chamber_section}.
As shown in the figure, we can identify a low-density dust cavity surrounded by the dust shell.
The estimated diameters in the heliocentric galactic cartesian coordinates (X, Y, and Z directions; see Figure \ref{fig:Perseus_chamber} for the definition of X, Y, and Z) are
$D_X \simeq 156$ pc, $D_Y \simeq 110$ pc, and $D_Z \simeq 105$ pc, centering at $X \simeq -220$ pc, $Y \simeq 38$ pc, and $Z \simeq -80$ pc.
The ellipsoid's estimated shape is also shown in Figure \ref{fig:Perseus_chamber}, and the outline of the ellipsoid projected on the POS is shown in Figure \ref{fig:TAP40} (bottom panel).
The Taurus molecular cloud is located in front of this cavity, and the Perseus molecular cloud is located behind it.
Thus, we conclude that the two-component \emph{B}-fields observed in the Perseus cloud's direction at distances of 150 pc and 300 pc show the \emph{B}-field structure in front of and behind the dust cavity, respectively.

The H\small{I} shell and dust cavity are thought to be formed by the Per OB2 association.
The velocity field that is consistent with the assumed expanding motion of the shell was found for the Perseus molecular cloud in the H\small{I} \citep{1974IAUS...60..115S} and CO \citep{2006AJ....131.2921R,2006A&A...451..539S} line emissions (see, e.g., Figure 10 of \citealp{2006A&A...451..539S}).
\citet{2018A&A...614A.100T} found that the LOS \emph{B}-field directions are different from each other in the north and south of the east-west extending Perseus molecular cloud.
This LOS \emph{B}-field is consistent with a scenario where the environment has impacted and influenced the field lines to form a bow-shaped magnetic field morphology as explored in \citet{1997ApJS..111..245H} and \citet[][also see \citealp{2018PASJ...70S..53I}]{2019A&A...632A..68T}.
$\theta_\mathrm{star}$ of the Perseus cloud \emph{B}-field component is $-30.0^\circ \pm 25.2^\circ$ (Section \ref{sec:qu_separation} and Table \ref{tab:p_efficiency}).
In the galactic coordinates, $\theta_\mathrm{star}^\mathrm{GAL}$ (measured from Galactic North to Galactic East) of the Perseus cloud \emph{B}-field is $-68.5^\circ \pm 24.9^\circ$.
This $\theta_\mathrm{star}^\mathrm{GAL}$ is nearly parallel to the galactic plane (see Figure \ref{fig:AngPbreak}, top panel), i.e., it is consistent with the global \emph{B}-field component of the galaxy \citep[see, e.g.,][Figure 9]{2012ApJ...757...14J}. This $\theta_\mathrm{star}^\mathrm{GAL}$ is, therefore, in accordance with the formation scenario of the Perseus molecular cloud described above.
The impact of feedback on the \emph{B}-fields is explored on a forthcoming paper (Tahani et al., in prep).

On the other hand, the orientation of the Taurus \emph{B}-field is close to perpendicular to the galactic plane ($\theta_\mathrm{star}^\mathrm{GAL} = 28.7^\circ \pm 19.8^\circ$; Figure \ref{fig:AngPbreak}, top panel).
It is difficult to form a \emph{B}-field structure close to perpendicular to the galactic plane (i.e., perpendicular to the global \emph{B}-field of the galaxy) if the ISM is compressed simply along the LOS.
Thus, the observed $\theta_\mathrm{star}$ in front of and behind the dust cavity is not compatible with the simple expansion of H\small{I} bubble in a uniform \emph{B}-field.

As shown in the discussion in this section, thanks to the advent of Gaia data, we can now isolate the \emph{B}-fields associated with each of the multiple clouds superimposed along the LOS and map them to the 3D structure of the ISM \citep[see also][and the references therein]{2019ApJ...872...56P}.
This isolation of the \emph{B}-fields is important in studying the  \emph{B}-field properties of individual clouds, e.g., applying the DCF method \citep{1951PhRv...81..890D,1953ApJ...118..113C} to estimate the \emph{B}-field strength.

\subsection{Consistency between the Optical and the Planck-observed \emph{B}-field}
\label{sec:PlanckCorrelation}

\begin{figure}[tp]
\centering
\includegraphics[width=\linewidth]{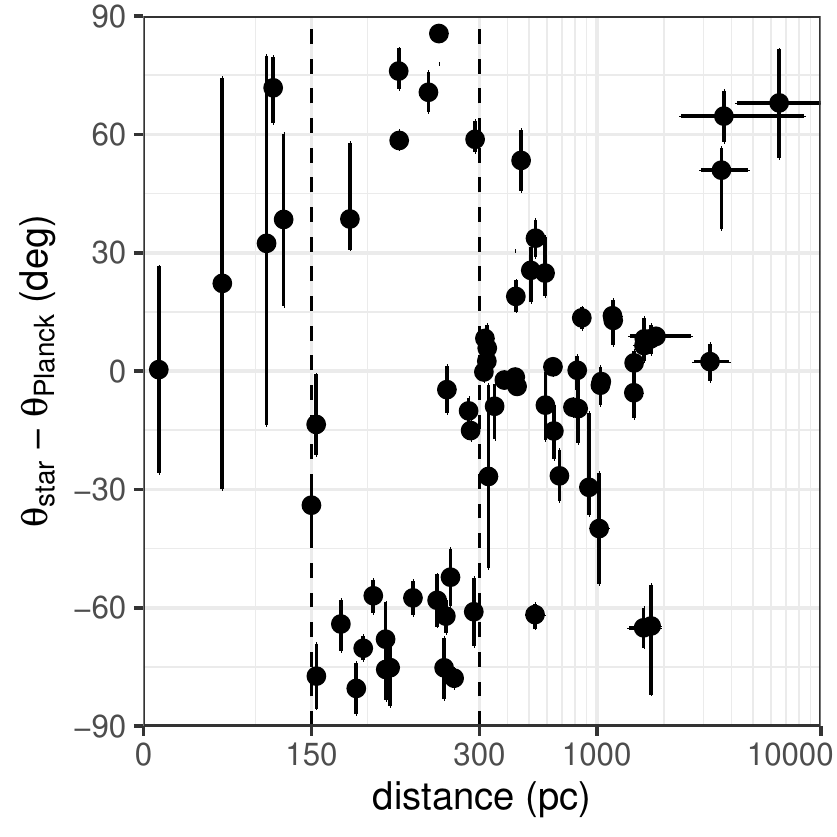}
\caption{
The angle difference between the optical polarimetry ($\theta_\mathrm{star}$) and the Planck-observed \emph{B}-field ($\theta_\mathrm{Planck}$) in the Perseus molecular cloud's direction.
The spatial resolution of the Planck observation is set as $10'$.
The offset angle is shown as a function of stellar distances observed by Gaia.
Note that the horizontal axis is a linear scale below 300 pc, but a logarithmic scale above 300 pc.
The dotted lines at $d = 150$ pc and 300 pc indicate the two breakpoints identified in $\theta_\mathrm{star}$ and $P$ (Section \ref{sec:breakpoints}), corresponding to the Taurus and Perseus clouds (Section \ref{sec: Taurus foreground}).
}
\label{fig:offset_distance}
\end{figure}

In Figure \ref{fig:Goodman}, we showed the spatial distribution of $\theta_\mathrm{star}$ and $\theta_\mathrm{Planck}$.
We note that the Group 1 $\theta_\mathrm{star}$ ($d > 300$ pc; red points in Figure \ref{fig:Goodman}) has a good one-to-one correspondence with $\theta_\mathrm{Planck}$ in the same position, despite their spatial variations.
Figure \ref{fig:offset_distance} shows the offset angle between $\theta_\mathrm{star}$ and $\theta_\mathrm{Planck}$ as a function of stellar distance.
For stars closer than 300 pc, the offset angle is $\theta_\mathrm{star} - \theta_\mathrm{Planck} = -71^\circ \pm 45^\circ$ and shows large variation, while for stars farther than 300 pc, the offset angle is $\theta_\mathrm{star} - \theta_\mathrm{Planck} = 2^\circ \pm 28^\circ$.
$\theta_\mathrm{Planck}$ traces the \emph{B}-field approximately in proportion to the column density along the LOS.
Thus, $\theta_\mathrm{Planck}$ mainly traces the Perseus molecular cloud, with an additional contribution from the foreground cloud (see Figure \ref{fig:cloud_distance_profile}).
As discussed in Section \ref{sec:qu_separation}, this also applies to $\theta_\mathrm{star}$ of Group 1.
As a result, $\theta_\mathrm{Planck}$ agrees well with the Group 1 $\theta_\mathrm{star}$ as they trace the similar ISM on the LOS.
On the other hand, $\theta_\mathrm{star}$ of Group 2 and Group 3 has no contribution from the Perseus cloud, and the foreground contribution traced by them is much smaller than that of the Perseus cloud in the background.
Therefore, $\theta_\mathrm{star}$ of Group 2 and Group 3 does not correlate with $\theta_\mathrm{Planck}$.
Thus, we conclude that $\theta_\mathrm{star}$ and $\theta_\mathrm{Planck}$ observations of the Perseus molecular cloud's direction with stellar distances $d > 300$ pc are in good agreement with each other, despite the fact that their spatial resolutions are largely different from each other.
The spatial resolution of $\theta_\mathrm{Planck}$ is $10'$, which corresponds to $\sim 1$ pc at 300 pc from the sun.
On the other hand, the spatial resolution of $\theta_\mathrm{star}$ is the size of stellar photospheres, which is much smaller than the Planck's beam size.

The good one-to-one correspondence between $\theta_\mathrm{star}$ and $\theta_\mathrm{Planck}$ was first pointed out by \citet[][see also \citealp{2019ApJ...871L..15G}]{2015A&A...576A.106P}.
\citet{2016A&A...596A..93S} further analyzed the $\theta_\mathrm{star}$--$\theta_\mathrm{Planck}$ correlation and investigated possible small-scale structures of the \emph{B}-field inside the Planck beam.
They concluded that the \emph{B}-field inside the Planck beam is uniform or randomly fluctuated around the mean $\theta$, because $\theta_\mathrm{Planck}$ agrees with $\theta_\mathrm{star}$.
Our observed good correlation between $\theta_\mathrm{Planck}$ and $\theta_\mathrm{star}$ supports their conclusion.
\begin{figure}[tp]
\centering
\includegraphics[width=\linewidth]{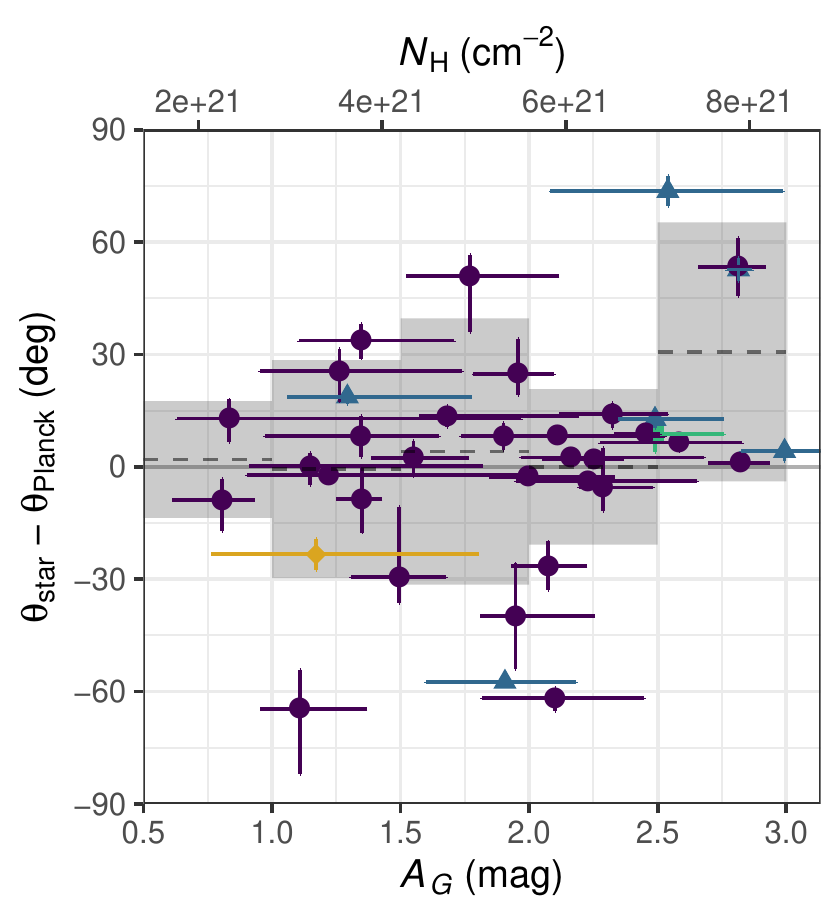}
\caption{
The angle difference between the optical polarimetry ($\theta_\mathrm{star}$) and the Planck-observed \emph{B}-field ($\theta_\mathrm{Planck}$) in the Perseus molecular cloud's direction.
The spatial resolution of the Planck observation is set as $10'$.
The offset angle is shown as a function of stellar extinction $A_G$ observed by Gaia.
Corresponding $N_\mathrm{H}$ values are also shown at the top axis of the figure, assuming $N_\mathrm{H} = A_G \cdot 2.21\times 10^{21} / 0.789$ \citep{2009MNRAS.400.2050G,2019ApJ...877..116W}.
The colors of the points are the same as in Figure \ref{fig:AngPbreak}.
The dashed lines and the shaded area indicate the circular mean angle differences per $A_G = 0.5$ mag and their standard deviations.
}
\label{fig:offset_AG}
\end{figure}
We show the offset angle between $\theta_\mathrm{star}$ and $\theta_\mathrm{Planck}$ as a function of the stellar extinction in Figure \ref{fig:offset_AG}.
The data show good agreement (angle differences consistent with zero) with no clear dependence on column density.
The data shown here suggest that the small scale \emph{B}-field is smooth in the column density range up to $A_G \sim 3$ mag or hydrogen column density $N_\mathrm{H} \sim 10^{22}~(\mathrm{cm}^{-2})$.

In addition, Figure \ref{fig:offset_distance} suggests that the correlation between $\theta_\mathrm{Planck}$ and $\theta_\mathrm{star}$ doesn't change with the stellar distance beyond 300 pc.
As shown in Section \ref{sec:breakpoints} and Figure \ref{fig:AngPbreak}, $\theta_\mathrm{star}$ and $P_\mathrm{star}$ show no clear variation as a function of the stellar distance.
The good correlation between $\theta_\mathrm{star}$ and $\theta_\mathrm{Planck}$ thus indicate that the ISM between and behind the clouds does not significantly contribute to polarization in either emission ($\theta_\mathrm{Planck}$) or extinction ($\theta_\mathrm{star}$).
In other words, the polarization is produced only in dense clouds along the LOS.

\subsection{Small-scale \emph{B}-field in low- and high-column density regions}
\label{sec:small_scale}

\citet{2020ApJ...899...28D} performed sub-mm polarimetry observations of NGC 1333, a dense star-forming region in the Perseus molecular cloud.
They observed the region with POL-2 on JCMT, and revealed the \emph{B}-field orientation ($\theta_\mathrm{JCMT}$) with a high spatial resolution of $\simeq 0.02$ pc.
The observed region has ISM column density $N_\mathrm{H} \gg 10^{23}~\mathrm{(cm^{-2}}$; also see Figure \ref{fig:Goodman}).
In contrast to the good agreement between $\theta_\mathrm{star}$ and $\theta_\mathrm{Planck}$ described in the previous section, $\theta_\mathrm{JCMT}$ is not well correlated with $\theta_\mathrm{Planck}$.
Instead, $\theta_\mathrm{JCMT}$ shows significantly more complex structure on $< 0.5$ pc scales.
\citet{2020ApJ...899...28D} found that the small-scale \emph{B}-field appears to be twisted perpendicular to the gravitationally super-critical massive filaments, most probably due to the dense filaments' formation.

Hence, the small scale ($< 1$ pc) \emph{B}-field appears to be highly complicated for the regions with $N_\mathrm{H} \gg 10^{23}~(\mathrm{cm}^{-2})$ but smooth for the regions with $N_\mathrm{H} < 10^{22}~(\mathrm{cm}^{-2})$.
It is unknown about the small-scale \emph{B}-field structure for the regions with $N_\mathrm{H} = 10^{22}$ -- $10^{23}~(\mathrm{cm}^{-2})$.
This is because optical and NIR stellar polarimetry cannot trace $N_\mathrm{H} \gg 10^{22}~(\mathrm{cm}^{-2})$ ISM, while ground-based and airborne sub-mm telescopes currently lack sensitivity to trace $N_\mathrm{H} \ll 10^{24}~(\mathrm{cm}^{-2})$ ISM.
But as described below, observations of the LOS \emph{B}-field strength and Planck polarimetry with the lower spatial resolution hint at the increasing complexity of small-scale \emph{B}-field structure at $N_\mathrm{H} > 10^{22}~(\mathrm{cm}^{-2})$.
\citet{2012ARA&A..50...29C} showed that the LOS \emph{B}-field strength measured by Zeeman splitting measurements show a general increase with an increasing $N_\mathrm{H}$ when $N_\mathrm{H} > 10^{22}~(\mathrm{cm}^{-2})$.
In such high column density region, \citet{2015A&A...576A.104P} observed a sharp drop in $P_\mathrm{Planck}$ where $N_\mathrm{H} \geq 1.5 \times 10^{22}$ (cm$^{-2}$).
They also reported an increase of the polarization angle dispersion at the same column density range.
\citet{2016A&A...586A.138P} and \citet{2019A&A...629A..96S} found that the relative orientation of the Planck-observed \emph{B}-fields to the long axes of dense filaments changes systematically from being parallel to perpendicular at $N_\mathrm{H} \approx 5 \times 10^{21}$ (cm$^{-2}$), which is consistent with the result reported by \citet{2020ApJ...899...28D}.

Photometric observations also suggest that ISM clouds make a significant change in this column density range.
\citet{1998ApJ...502..296O} found recently formed protostars only in regions with $N_\mathrm{H_2} > 8 \times 10^{21}$ (cm$^{-2}$) in the Taurus molecular cloud.
\citet{2004ApJ...611L..45J} and \citet{2006ApJ...646.1009K} found no obvious substructures below an $N_\mathrm{H} \sim 10^{22}~(\mathrm{cm}^{-2})$ in their sub-millimeter continuum observations of the Ophiuchus cloud and the Perseus cloud, respectively.
Many authors have pointed out that there is a fiducial threshold of $N_\mathrm{H} \simeq 0.6$ -- $2 \times 10^{22}$ (cm$^{-2}$) for the star formation \citep[e.g.,][]{2010ApJ...724..687L,2012ApJ...745..190L,2010ApJ...723.1019H,2014ApJ...782..114E,2015A&A...584A..91K,2016MNRAS.459..342M,2019A&A...622A..52Z}, though some counterexample are reported by, e.g., \citet[][lower threshold value]{2020ApJ...904..172D} and \citet[][no threshold value]{2020ApJ...896...60P}.
Interestingly, the critical line mass of ISM filaments ($M_\mathrm{line,crit} = 2c_s^2/G \simeq 16~M_\odot~\mathrm{pc^{-1}}$ for $T_\mathrm{gas}=10$ K; \citealp{1963AcA....13...30S}; \citealp{1964ApJ...140.1056O}; \citealp{1997ApJ...480..681I}) is consistent with this threshold $N_\mathrm{H}$ value if we adopt the typical 0.1 pc width of the filaments \citep{2014prpl.conf...27A}.

These observations described above suggest that ISM clouds make a critical change above $N_\mathrm{H} \sim 10^{22}~(\mathrm{cm}^{-2})$.
Suppose the small scale \emph{B}-field becomes complex and the \emph{B}-field strength increases at $N_\mathrm{H} > 10^{22}~(\mathrm{cm}^{-2})$.
In that case, it could potentially show that the molecular clouds become gravitationally supercritical when $N_\mathrm{H} \gtrsim 10^{22}~(\mathrm{cm}^{-2})$, and thus the small scale structure in the molecular clouds are formed in this column density range.
The \emph{B}-field might be bent and distorted in the formation of gravitationally supercritical small scale structures or dense filaments.
Therefore, the \emph{B}-field structure may record these small-scale structures' formation history.
High spatial resolution (e.g., $< 15\arcsec$) interstellar \emph{B}-field observations in this column density range, if achieved, could provide important information on the formation of cloud structure in the early stages of star formation.

\section{Conclusions}
\label{sec:conclusion}

We studied the optical polarimetry data in the Perseus molecular cloud's direction together with the Gaia parallax distance of each star.
We found that observed values of both polarization angles and fractions show discrete jump at 150 pc and 300 pc and otherwise remain constant.
Thus, the polarization is originated in the dust clouds at 150 pc and 300 pc. The ISM between and behind the clouds does not make a significant contribution to the polarization.

The dust cloud at 300 pc corresponds to the Perseus molecular cloud.
We estimate the POS \emph{B}-field orientation angle of the Perseus molecular cloud as $-30.0^\circ \pm 25.2^\circ$.
The Perseus cloud has the highest column density in LOS, and thus Planck observations of the dust continuum in this direction mainly trace this cloud.
The optical polarization angles (stellar distances $d > 300$ pc) and the Planck-observed \emph{B}-field orientations are therefore well aligned with each other, although their spatial resolutions are largely different.

The dust cloud at a distance of 150 pc is faint with $A_G < 1$ mag, and the POS \emph{B}-field orientation angle is $+66.8^\circ \pm 19.1^\circ$.
We identified this cloud as the outer edge of the Taurus molecular cloud at the same distance.

The two dust clouds are at the front- and back-sides of a dust cavity, which corresponds to an H\small{I} bubble structure generally associated with the Per OB2 association.
We estimate the size of the dust cavity as 100 -- 160 pc.
The observed two components of the POS \emph{B}-field orientations show the \emph{B}-field orientations in front of and behind the cavity, which are nearly perpendicular to one another.

Thanks to the advent of Gaia data, we can now isolate the \emph{B}-fields associated with each of the multiple clouds superimposed along the LOS and map them to the 3D structure of the ISM.
This process can be applied to other regions, and we hope such efforts provide an important step toward understanding the 3D \emph{B}-field structure of the ISM.


\acknowledgments


This research has been supported by Grants-in-Aid for Scientific Research (25247016, 18H01250, 19H01938) from the Japan Society for the Promotion of Science.
This research was supported in part at the SOFIA Science Center, which is operated by the Universities Space Research Association under contract NNA17BF53C with the National Aeronautics and Space Administration.
M.M. is supported by JSPS KAKENHI grant No. 20K03276.
S.I.S acknowledges support from the Natural Science and Engineering Research Council of Canada (NSERC), RGPIN-2020-03981.
C.L.H.H. acknowledges the support of the NAOJ Fellowship and JSPS KAKENHI grants 18K13586 and 20K14527.
D.J. and R.P. are supported by the National Research Council  of  Canada  and  by  a  Natural  Sciences  and  Engineering  Research  Council  of  Canada  (NSERC)  Discovery Grant. 
J.K.is supported by JSPS KAKENHI grant No. 19K14775.
M.T. is supported by JSPS KAKENHI grant Nos. 18H05442, 15H02063, and 22000005.
This work has made use of data from the European Space Agency (ESA) mission
{\it Gaia} (\url{https://www.cosmos.esa.int/gaia}), processed by the {\it Gaia}
Data Processing and Analysis Consortium (DPAC,
\url{https://www.cosmos.esa.int/web/gaia/dpac/consortium}). Funding for the DPAC
has been provided by national institutions, in particular the institutions
participating in the {\it Gaia} Multilateral Agreement.
This work has make use of observations obtained with Planck (http://www.esa.int/Planck), an ESA science mission with instruments and contributions directly funded by ESA Member States, NASA, and Canada.
This research has also made use of the SIMBAD database and of NASA's Astrophysics Data System Bibliographic Services.

\facilities{Gaia, Planck}
\software{strucchange \citep{strucchange,breakpoints}, astropy \citep{2013A&A...558A..33A}}





\appendix

\section{Data List}
\label{sec:used_data}

We list the cross-matching results between stellar polarimetry data and the Gaia catalog in Table \ref{tab:stellar_data}.
See Section \ref{sec:DR2} for the cross-matching procedure.

\begin{table*}
\begin{center}
\caption{Data Identification}
\label{tab:stellar_data}
\begin{splittabular}{crccccrrBccrrcrrcc}
\hline
\hline
Ref.$^\mathrm{a}$ & No.$^\mathrm{b}$ & R.A. (ICRS) & Dec. (ICRS) & $P$ & $\delta P$ & \multicolumn{1}{c}{$\theta$} & \multicolumn{1}{c}{$\delta \theta$} & Gaia No.$^\mathrm{c}$ & distance & \multicolumn{2}{c}{$\delta$distance} & $A_G$ & \multicolumn{2}{c}{$\delta A_G$} & RUWE$^\mathrm{d}$ & Bad$^\mathrm{e}$\\
& & (deg) & (deg) & (\%) & (\%) & (deg) & (deg) & & \multicolumn{1}{c}{(pc)} & (pc) & (pc) & (mag) & (mag) & (mag) & & \\
\hline
1 & 1 & 50.89583 & 30.54518 & 1.15 & 0.10 & 80 & 2 & 123890260094849536 & 292.1 & +5.7 & -5.5 & \multicolumn{1}{c}{---} & \multicolumn{1}{c}{---} & \multicolumn{1}{c}{---} & 0.87 & 0\\
1 & 2 & 50.96720 & 30.58532 & 0.78 & 0.09 & 82 & 3 & 123891256527257088 & 290.2 & +5.4 & -5.3 & 0.3250 & +0.1290 & -0.1534 & 0.93 & 0\\
1 & 3 & 51.07634 & 30.26754 & 0.02 & 0.06 & -35 & 6 & 120877357716054528 & 115.8 & +0.9 & -0.8 & 0.1053 & +0.1098 & -0.0883 & 0.98 & 0\\
1 & 4 & 51.33642 & 30.06133 & 0.13 & 0.10 & -74 & 21 & 120858764801761152 & 125.1 & +0.8 & -0.8 & 0.2540 & +0.2108 & -0.2150 & 0.99 & 0\\
1 & 5 & 51.44488 & 30.74159 & 0.64 & 0.09 & 76 & 4 & 120987682540431744 & \multicolumn{1}{c}{---} & \multicolumn{1}{c}{---} & \multicolumn{1}{c}{---} & \multicolumn{1}{c}{---} & \multicolumn{1}{c}{---} & \multicolumn{1}{c}{---} & 1.07 & 1\\
1 & 6 & 51.45880 & 30.93167 & 1.38 & 0.09 & -70 & 2 & 123999730221048320 & 313.7 & +6.9 & -6.6 & \multicolumn{1}{c}{---} & \multicolumn{1}{c}{---} & \multicolumn{1}{c}{---} & 0.98 & 0\\
1 & 7 & 51.61334 & 30.15060 & 1.14 & 0.07 & 85 & 2 & 120819186679044480 & 296.1 & +5.6 & -5.4 & 0.7603 & +0.4418 & -0.4440 & 1.03 & 0\\
1 & 8 & 51.62439 & 30.78881 & 1.97 & 0.09 & -59 & 1 & 120991672565773568 & 818.7 & +53.2 & -47.1 & 1.1480 & +0.6735 & -0.2385 & 1.11 & 0\\
1 & 9 & 51.82006 & 30.03606 & 0.55 & 0.09 & 76 & 5 & 120803686141726336 & 254.4 & +5.0 & -4.8 & \multicolumn{1}{c}{---} & \multicolumn{1}{c}{---} & \multicolumn{1}{c}{---} & 0.95 & 0\\
1 & 10 & 51.87204 & 30.72097 & 1.33 & 0.05 & -61 & 1 & 120979406139173376 & 350.3 & +9.3 & -8.9 & 0.8060 & +0.1260 & -0.1941 & 1.09 & 0\\
\hline
\end{splittabular}
\end{center}
{\bf Notes.}\\
$^\mathrm{a}$ Reference number. 1: Optical \citep[][Table 3]{1990ApJ...359..363G}, 2: $R$-band \citep[][Table 5]{2011AJ....142...33A}, 3: $J$-band \citep[][Table 5]{2011AJ....142...33A}, 4: $K$-band \citep[][Table 2]{1988MNRAS.231..445T}.\\
$^\mathrm{b}$ Source number in each reference.\\
$^\mathrm{c}$ Gaia source ID in DR 2.\\
$^\mathrm{d}$ see Section \ref{sec:DR2} for the definition of RUWE.\\
$^\mathrm{e}$ Bad flag. Bad $=1$ indicates that the data entry is discarded from the analysis in this paper because of its large RUWE ($> 1.4$), non-reliable Gaia parallax (NA or negative values), or erroneous identification ($G$ magnitude $> 15$ mag for data by \citealp{1990ApJ...359..363G}).\\
(This table is available in its entirety in machine-readable form.
A portion is shown here for guidance regarding its form and content.)
\end{table*}


\bibliography{paper3}{}
\bibliographystyle{aasjournal}


\listofchanges

\end{document}